\documentclass[fleqn,usenatbib]{mnras}

\usepackage[T1]{fontenc}

\usepackage{graphicx}	% Including figure files
\usepackage[normalem]{ulem} %for \sout{}

\usepackage{amsmath}	% Advanced maths commands
\usepackage{amssymb}	% Extra maths symbols
\usepackage{multirow}
\usepackage{xcolor}

\newcommand{\strikethroughEC}{\bgroup\markoverwith{\textcolor{violet}{\rule[0.5ex]{2pt}{1pt}}}\ULon}

\newcommand{\soutPC}{\bgroup\markoverwith{\textcolor{cyan}{\rule[0.5ex]{2pt}{1pt}}}\ULon}

\newcommand{\soutEV}{\bgroup\markoverwith{\textcolor{magenta}{\rule[0.5ex]{2pt}{1pt}}}\ULon}

\DeclareRobustCommand{\VAN}[3]{#2}
\let\VANthebibliography\thebibliography
\def\thebibliography{\DeclareRobustCommand{\VAN}[3]{##3}\VANthebibliography}
\DeclareRobustCommand{\DE}[3]{#2}
\let\DEthebibliography\thebibliography
\def\thebibliography{\DeclareRobustCommand{\DE}[3]{##3}\DEthebibliography}

\newcommand{\placetabAGAMA}{
\begin{table*}
\caption{The main {\textsc{Agama}} parameters to set up the FCC\,170 $N$-body model.}             
\centering
\resizebox{\textwidth}{!}{
\begin{tabular}{c c c c c c c c}     % 7 columns 
\hline\hline       
Particle type & Component & Density profile & DF family & Total mass & Scale radius & Scale height & Particle mass \\
 &  &  &  & [$10^{9}$~M$_{\sun}$] & [kpc] & [kpc] & [M$_{\sun}$] \\ 
(1) & (2) & (3) & (4) & (5) & (6) & (7) & (8) \\
\hline \hline 

Black hole & Central SMBH & Plummer & - & $3.11\times10^{-2}$ & 0.001 & \ldots & $3.11\times10^{7}$ \\ 
\hline 

Dark matter & DM halo & Spheroid & QuasiSpherical & $2.48\times10^{3}$ & 24.3 & \ldots & $2.48\times10^{5}$ \\ 
\hline 

\multirow{4}{*}{Stars} & Bulge & S\'ersic & DoublePowerLaw & $1.25\times10^{1}$ & 0.5 & \ldots & \multirow{4}{*}{$7.32\times10^{3}$} \\
& Thin disc & Disk & QuasiIsothermal & 3.60 & 2.1 & 0.18 &  \\
& Thick disc & Disk & QuasiIsothermal & 9.87 & 2.15 & 0.43 & \\
& NSD & Disk & QuasiIsothermal & $9.11\times10^{-1}$ & 0.05 & 0.02 &  \\   
\hline   

\end{tabular}
}
\label{tab:AGAMA_params}
\begin{minipage}{\textwidth}
{\em Notes.} Col. (1): particle type. Col. (2): structural component. Col. (3): initial density profile. Col. (4): DF type. Col. (5): total mass of the component as the sum of the mass of all the individual particles. Col. (6): scale radius defined as the distance where the density profile drops down by a factor $e$. Col. (7): scale height of the disc density profile. Col. (8): mass of each particle.
\end{minipage}
\end{table*}
}

\newcommand{\placefigThinDiscHeating}{
\begin{figure*}
    \centering
    \includegraphics[scale=0.55]{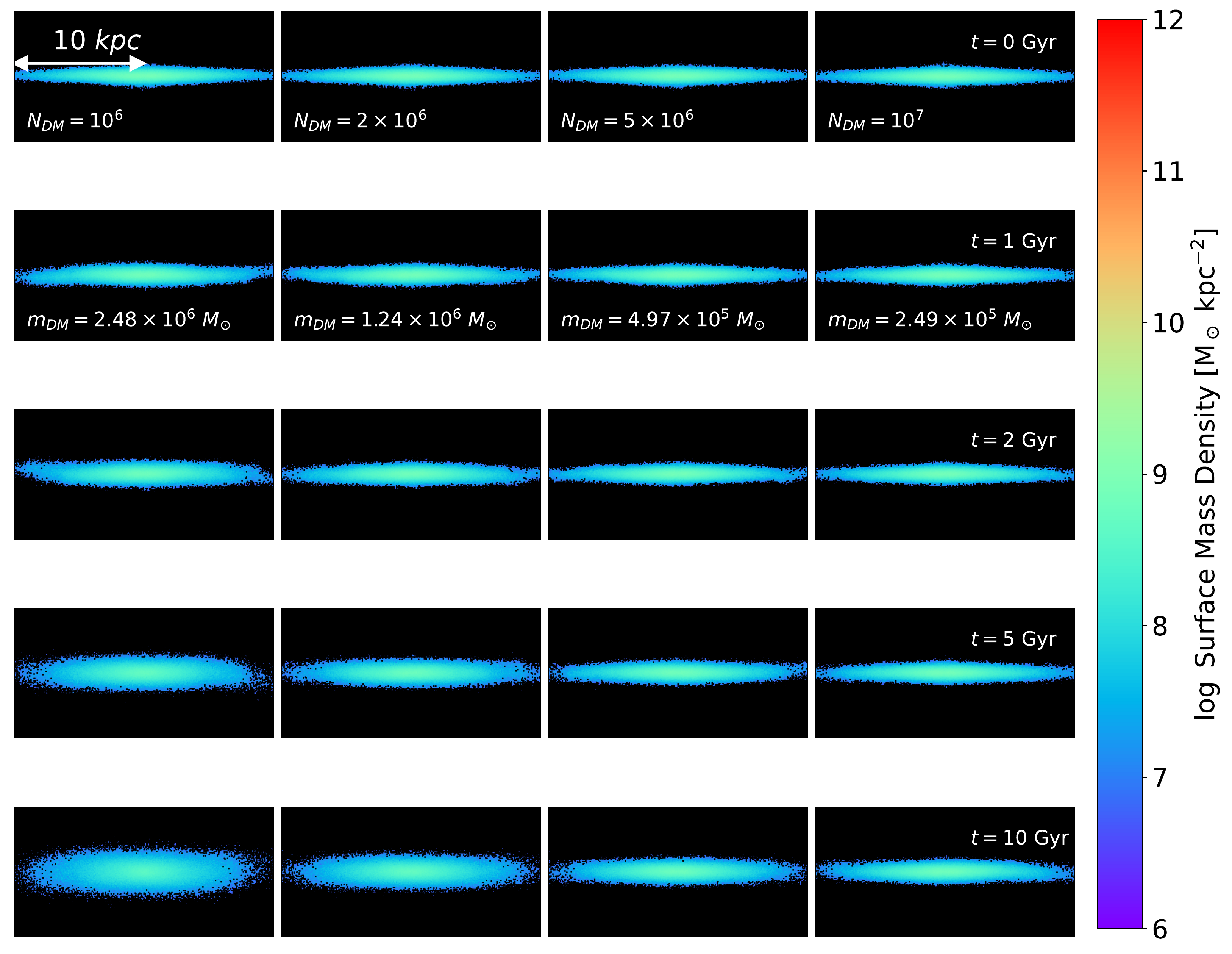}
    \caption{Surface mass density maps of the thin-disc particles of the FCC\,170 $N$-body model seen edge on at different times. Each column corresponds to a run with a different particle number and mass of DM particles, as given in the first and second row, respectively. Each panel is 20~kpc wide.
    }
    \label{fig:thin-disc_comparison}
\end{figure*}
}

\newcommand{\placefigNSDDiscHeating}{
\begin{figure*}
    \centering
    \includegraphics[scale=0.55]{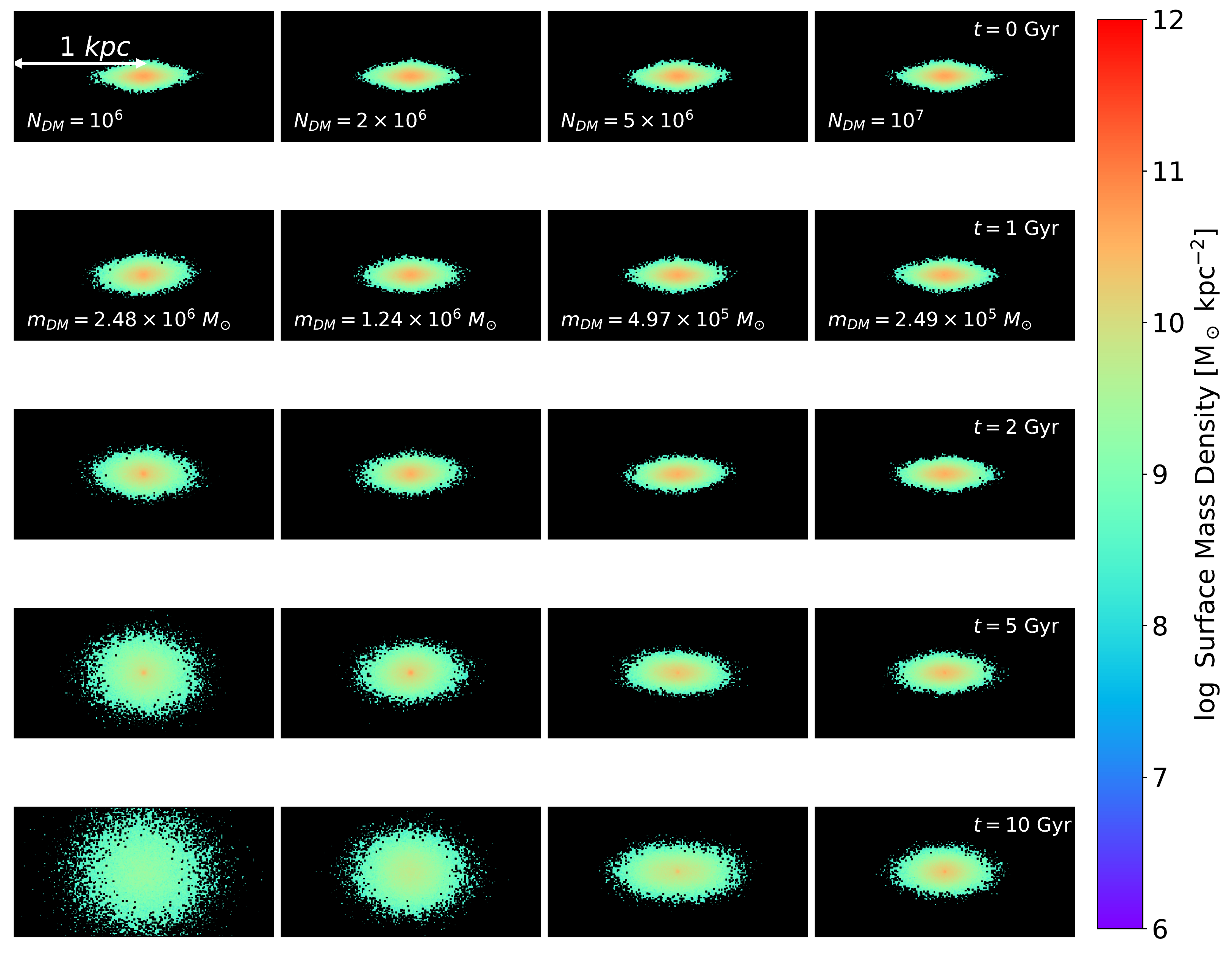}
    \caption{Same as Figure~\ref{fig:thin-disc_comparison}, but for the NSD particles.}
    \label{fig:NSD_comparison}
\end{figure*}
}

\newcommand{\placefigNSDprofiles}{
\begin{figure}
    \centering
    \includegraphics[scale=0.35]{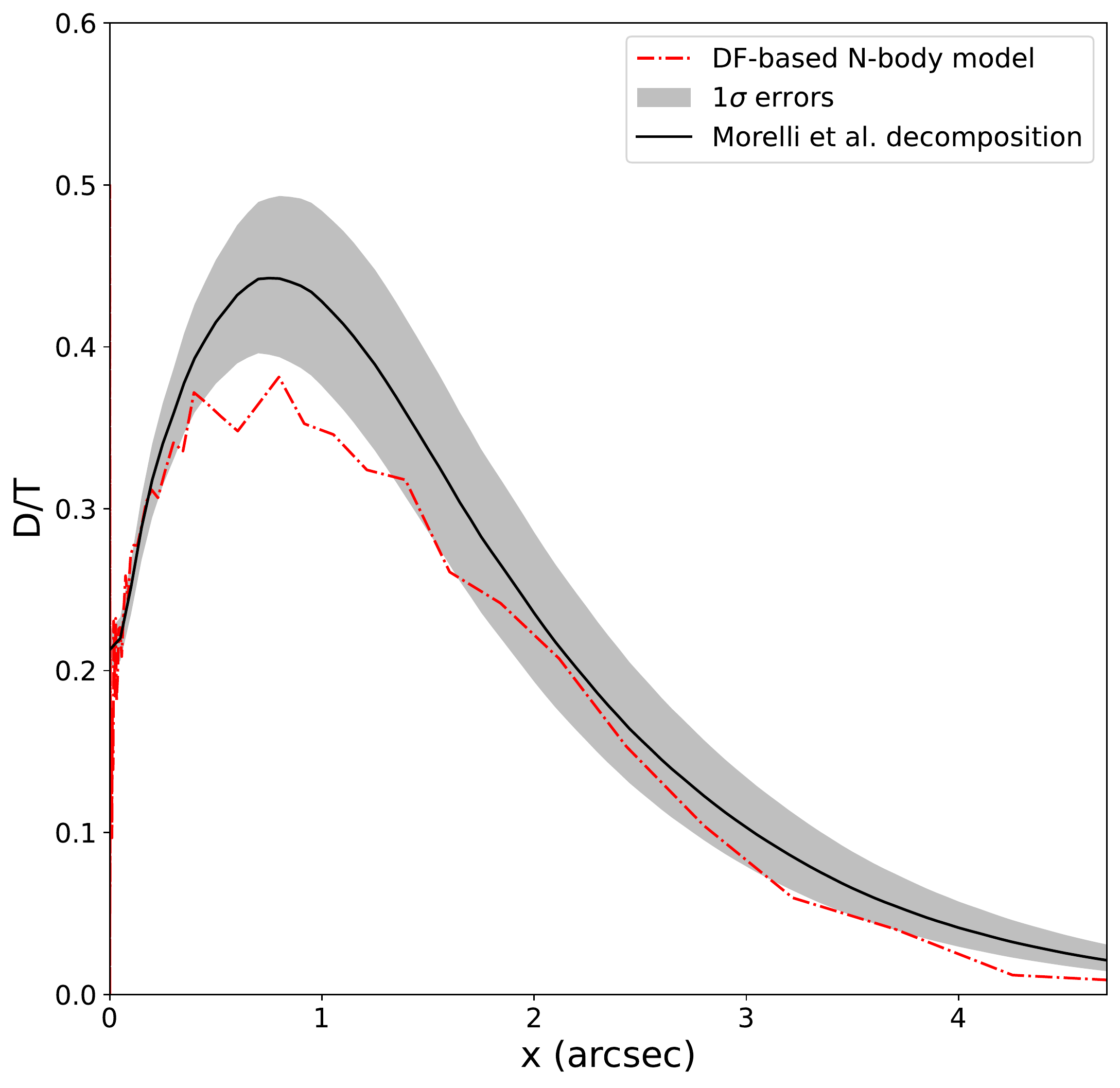}
    \caption{Radial profile of the disc-to-total fraction of the NSD in FCC\,170 along the major axis of the galaxy from the photometric decomposition by \textcolor{blue}{Morelli et al. (in prep.)} (black solid line) and our $N$-body model (red dash-dotted line) along with the $1\sigma$ errors for the photometric decomposition.}
    \label{fig:NSD_profile}
\end{figure}
}

\newcommand{\placefigNSDanalytic}{
\begin{figure*}
    \centering
    \includegraphics[scale=0.25]{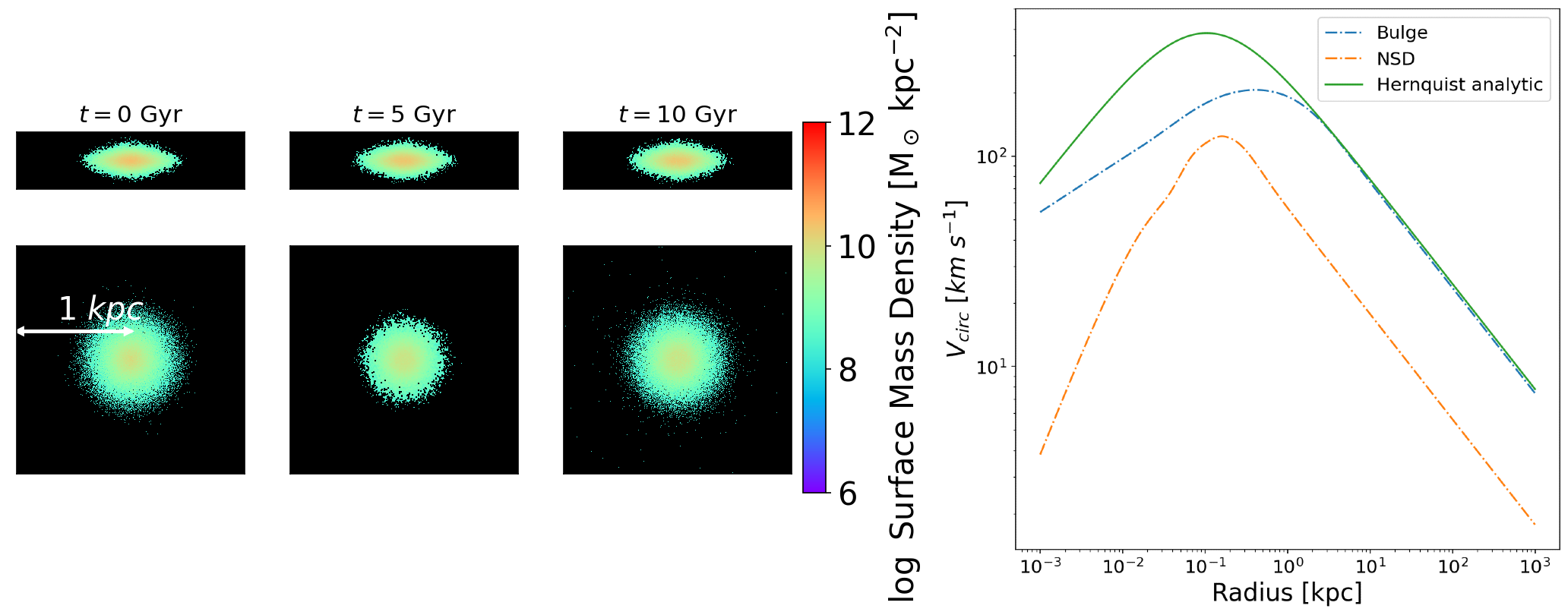}
    \caption{Left-hand panels: surface mass density maps of the NSD of the FCC\,170 $N$-body model at different times, where all the other structural components are replaced by an analytic \citet{Hernquist1990} mass model. Each column shows the edge-on (top panel) and face-on (bottom panel) view of the NSD particles at a given time. Each panel is 2~kpc wide. Right-hand panel: circular velocity of the bulge (blue dash-dotted line), NSD (orange dash-dotted line), and analytic Hernquist profile (green solid line) at different radii.
    }
    \label{fig:NSD_analytic}
\end{figure*}
}

\newcommand{\placefigFCCmodel}{
\begin{figure*}
    \centering
    \includegraphics[scale=0.45]{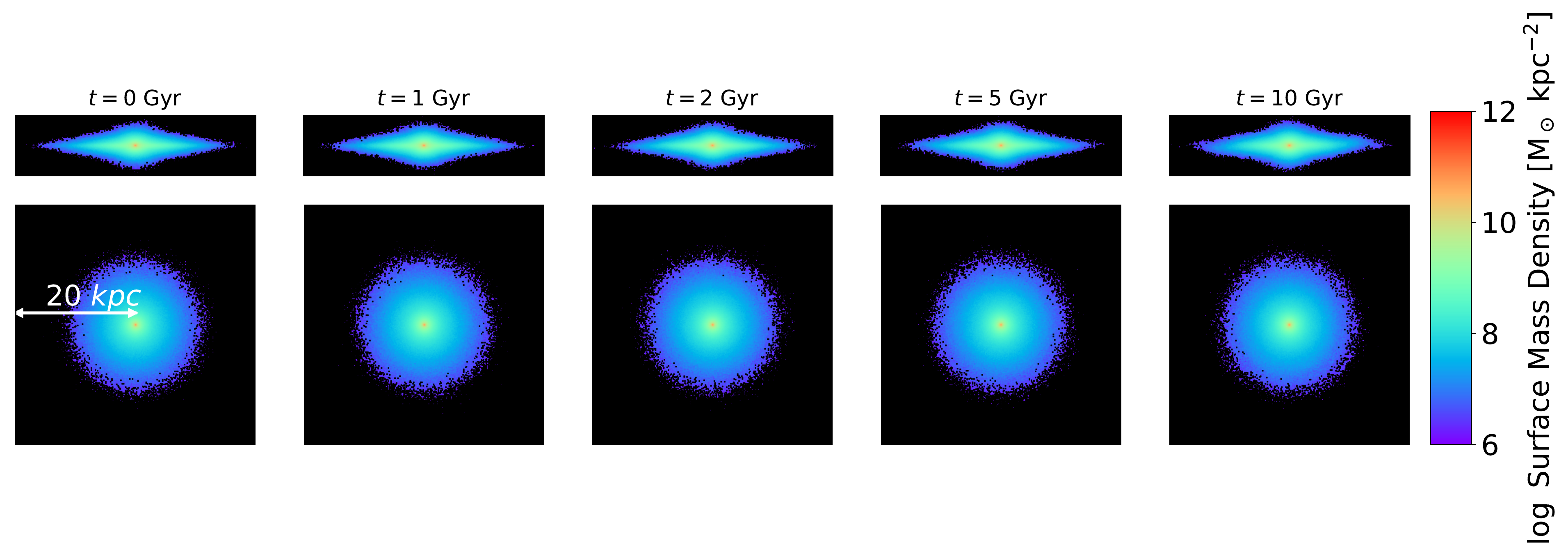}
    \caption{Surface mass density maps of the FCC\,170 $N$-body model at different times. Each column shows the edge-on (top panel) and face-on (bottom panel) view of the stellar component (i.e. NSD, bulge, and thin and thick discs) at a given time. Each panel is 40 kpc wide.
    }
    \label{fig:FCC170_model}
\end{figure*}
}

\newcommand{\placefigPhotometry}{
\begin{figure*}
    \centering
    \includegraphics[scale=0.6]{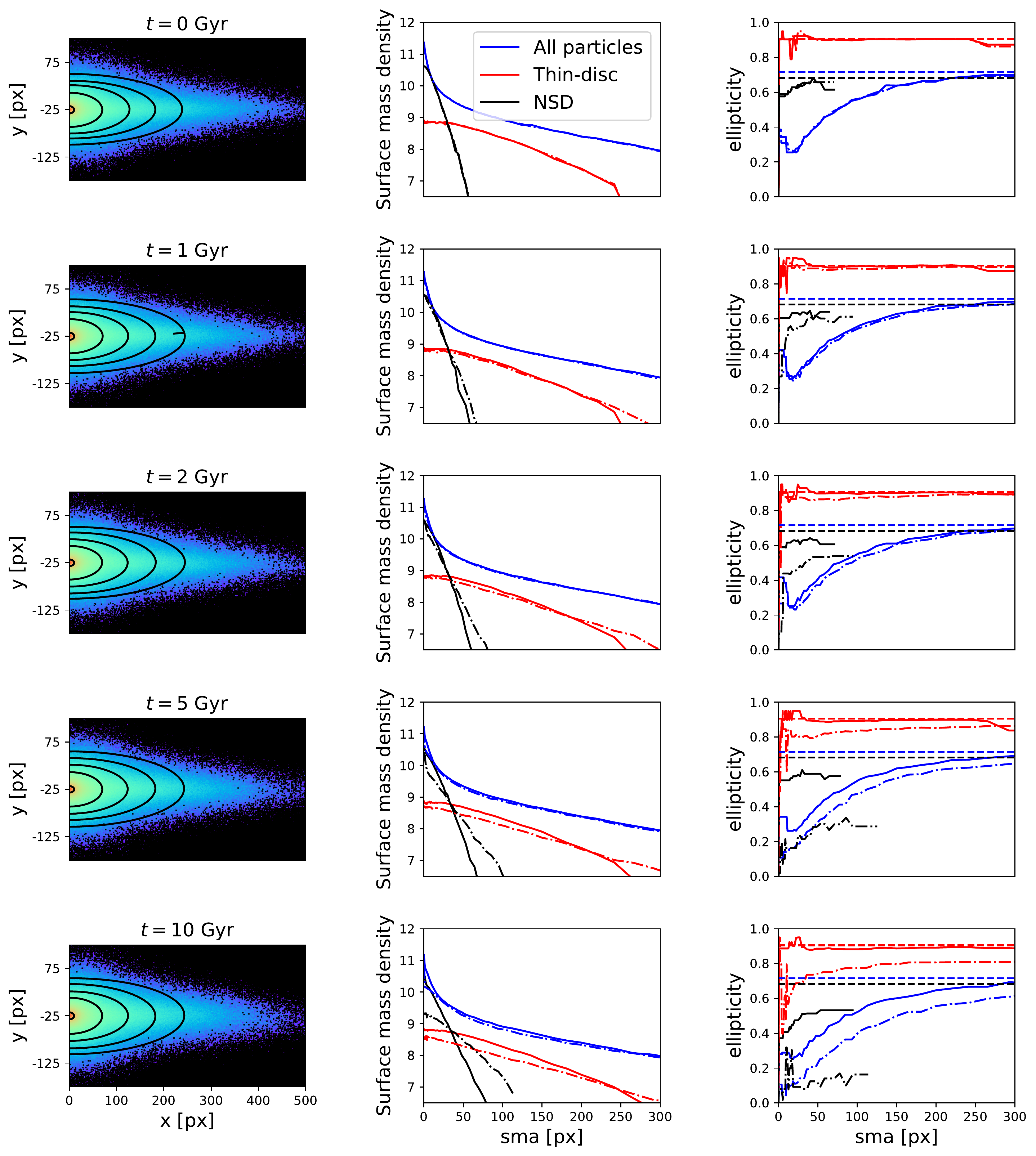}
    \caption{Photometric analysis of the FCC\,170 $N$-body model at different times. Left-hand column: edge-on view of the model along with some of the elliptical apertures adopted for the photometric analysis. Central column: surface mass density in arbitrary units along the major axis of the model for all stellar particles (blue lines), thin disc (red lines), and NSD (black lines). Right-hand column: ellipticity radial profiles along the major axis of the model for all stellar particles (blue lines), thin disc (red lines), and NSD (black lines). Dashed horizontal lines mark the value at which the ellipticity converges for each component at $t=0$. In the central and right-hand columns the solid and dot-dashed lines refer to the $N$-body models with $10^{7}$ and $10^{6}$ DM particles, respectively.
    }
    \label{fig:photometry}
\end{figure*}
}

\title[The fragility of thin discs in galaxies - I]{The fragility of thin discs in galaxies - I. Building tailored $N$-body galaxy models}

\author[P.~M. Gal\'an-de Anta et al.]{Pablo M. Gal\'an-de Anta,$^{1,2}$\thanks{E-mail: pgalandeanta01@qub.ac.uk}
Eugene Vasiliev,$^{3}$
Marc Sarzi,$^{2}$
Massimo Dotti,$^{4,5}$
Pedro~R. Capelo,$^{6}$
\newauthor
Andrea Incatasciato,$^{7}$
Lorenzo Posti,$^{8}$
Lorenzo Morelli$^{9}$
and Enrico Maria Corsini$^{10,11}$
\\
$^{1}$Astrophysics Research Centre, School of Mathematics and Physics, Queen's University Belfast, Belfast BT7 INN, UK\\
$^{2}$Armagh Observatory and Planetarium, College Hill, Armagh BT61 9DG, UK\\
$^{3}$Institute of Astronomy, Madingley road, Cambridge CB3 0HA, UK\\
$^{4}$Dipartimento di Fisica ``G. Occhialini'', Università degli Studi di Milano–Bicocca, Piazza della Scienza 3, I-20126 Milano, Italy\\
$^{5}$INFN, Sezione Milano–Bicocca, Piazza della Scienza 3, I-20126 Milano, Italy\\
$^{6}$Center for Theoretical Astrophysics and Cosmology, Institute for Computational Science, University of Zurich,\\ Winterthurerstrasse 190, CH-8057 Zürich, Switzerland\\
$^{7}$Institute for Astronomy, University of Edinburgh, Royal Observatory, Blackford Hill, Edinburgh EH9 3HJ, UK\\
$^{8}$Université de Strasbourg, CNRS UMR 7550, Observatoire astronomique de Strasbourg,\\ 11 rue de l'Université, F-67000 Strasbourg, France\\
$^{9}$Instituto de Astronomía y Ciencias Planetarias, Universidad de Atacama, Avenida Copayapu 485, Copiapó, Chile\\
$^{10}$Dipartimento di Fisica e Astronomia ``G. Galilei'', Università di Padova, Vicolo dell'Osservatorio 3, I-35122, Padova, Italy\\
$^{11}$INAF-Osservatorio Astronomico di Padova, Vicolo dell'Osservatorio 5, I-35122, Padova, Italy
}

\date{Accepted XXX. Received YYY; in original form ZZZ}

\pubyear{2022}

\begin{document}
\label{firstpage}
\pagerange{\pageref{firstpage}--\pageref{lastpage}}
\maketitle

\begin{abstract}
Thin stellar discs on both galactic and nuclear, sub-kpc scales are believed to be fragile structures that would be easily destroyed in major mergers. In turn, this makes the age-dating of their stellar populations a useful diagnostics for the assembly history of galaxies. We aim at carefully exploring the fragility of such stellar discs in intermediate- and low-mass encounters, using high-resolution $N$-body simulations of galaxy models with structural and kinematic properties tailored to actually observed galaxies. As a first but challenging step, we create a dynamical model of FCC\,170, a nearly edge-on galaxy in the Fornax cluster with multiple galactic components and including both a galactic scale and nuclear stellar disc (NSD), using detailed kinematic data from the Multi Unit Spectroscopic Explorer and a novel method for constructing distribution function-based self-consistent galaxy models. We then create $N$-body realisations of this model and demonstrate that it remains in equilibrium and preserves its properties over many Gyr, when evolved with a sufficiently high particle number. However, the NSD is more prone to numerical heating, which gradually increases its thickness by up to 22 per cent in 10~Gyr even in our highest-resolution runs. Nevertheless, these $N$-body models can serve as realistic representations of actual galaxies in merger simulations.
\end{abstract}
    

%%%%%%%%%%%%%%%%%%%%%%%%%%%%%%%%%%%%%%%%%%%%%%%%%%

%%%%%%%%%%%%%%%%% BODY OF PAPER %%%%%%%%%%%%%%%%%%

\section{Introduction}

About 70 per cent of the observed galaxies in the local Universe host (either thin or thick) kinematically cold disc-like structures in their stellar and gas distributions, regardless of the classification \citep{ilbert2006,weinmann2006,vandenbosch2007,choi2007,park2007,ledo2010}.  The high occurrence of disc galaxies and the paucity of past interaction signatures -- such as gas bridges or distinct rotating discs  \citep[e.g.][]{Corsini2014, Mazzilli2021} or extended stellar shells \citep[e.g.][]{Hernquist1988,Romanowsky2012,Pop2018,Onaka2018,Bilek2022} -- may appear at odds with the current hierarchical model of structure growth, in which minor and major mergers contribute significantly to the mass assembly of galaxies. Indeed, the effect of past interactions can dim in time, making necessary in-depth studies of the galaxy stellar populations to unveil their assembly history \citep{Davison2021,Mazzilli2021}. 

Kinematically cold thin discs are fragile structures subject to morphological transformations during interactions with their environment \citep{Vogelsberger2014a,Vogelsberger2014b,Genel2014,Sijacki2015,Joshi2020,Galan2022}, including galaxy mergers. For this reason, discs have been proposed as natural clocks of their last merger event \citep{Toomre1977,barnes1992,Hammer2009,Taranu2013,Deeley2017}.

It has however been recently observed that thin and kinematically cold structures could survive merger events spanning a wide range of mass ratios \citep[from 1:3 to <1:10;][]{Abadi2003,Robertson2006b,Purcell2009,Lotz2010,Moster2010}. In some cases, disc galaxies might even survive major encounters, leaving a prominent disc component in the merger remnant \citep[e.g.][]{Springel2005,Naab2006,Robertson2006b,Governato2007,capelo2015}. \citet{Athanassoula2016}, \citet{Sparre2017}, and \citet{Peschken2020} show that merger remnants of wet major mergers can induce the formation of galactic discs.

Alternative kinematically cold tracers of the last merger event are nuclear stellar discs (NSDs), originally unveiled in Hubble Space Telescope images \citep{Jaffe1994,Bosch1994}, where they appeared as razor-thin disc structures of a few hundred pc across lying at the centre of galaxies \citep{Pizzella2002}. \citet{ledo2010} presented a catalogue of NSDs in a wide variety of early-type galaxies getting a rough estimation for the number of NSDs residing in galactic nuclei to be about 20 per cent, making them a structure commonly present in the Universe. However, the properties of the stellar populations (i.e. age, metallicity, and star formation time-scale) have been derived only for a few NSDs \citep{Sarzi2016, Corsini2016}.

\citet{ledo2010} and \citet{sarzi2015} tested for the first time the fragility of NSDs against mergers, by performing a set of pure $N$-body simulations consisting of a NSD, a stellar halo, and a supermassive black hole (SMBH) in interaction with a secondary SMBH. In particular, \citet{sarzi2015} explored a broad region of the merger parameter space [circular orbits with different inclinations, impact parameters, and masses, including major (1:1), intermediate (1:5), and minor (1:10) mergers] and showed that these discs cannot survive any major encounter but can withstand minor ones. 

These first studies relied on rather idealised representations of the inner regions of the NSD host galaxies. More realistic studies modelling the whole stellar distribution of NSD hosts are needed in order to confirm the conclusions presented in \citet{sarzi2015}, allowing to consistently gauge the effect on the large-scale galactic disc and on the NSD.

To isolate the role of mergers and other physical processes in galaxy evolution, it is common to construct the initial conditions for the simulations as stationary equilibrium configurations. The methods for constructing these models can be grouped into a few categories. Jeans equations offer the fastest and least demanding approach, usually relying only on the first two moments of the stellar distribution function \citep[DF; e.g.][]{Cappellari2008,Mamon2013}, although the use of higher-order moments \citep[e.g.][]{Lokas2003,Richardson2013,Read2017} may help to lift the so-called mass--anisotropy degeneracy inherent to this approach \citep{Dejonghe1992}. It should be stressed that, while being fast and simple to produce, $N$-body realisations of galaxy models based on Jeans equations are not guaranteed to be in full equilibrium \citep{Kazantzidis2004}.

At the other end of the spectrum are made-to-measure $N$-body codes capable of guiding the model to meet specific observational constraints \citep{Syer1996,deLorenzi2007,Yurin2014}, which are very flexible but also expensive to run. The \citet{Schwarzschild1979} orbit-superposition method (in numerous implementations) has also been used to construct flexible models of observed galaxies \citep[e.g.][]{Cretton1999,Gebhardt2003,vdBosch2008} and generate the initial conditions for $N$-body simulations \citep[e.g.][]{Vasiliev2015}. Finally, self-consistent models based on DFs have been used both in the theoretical and simulation context \citep[e.g.][]{Kuijken1995,Debattista2000} and observational applications \citep[e.g.][]{Widrow2005,Piffl2015,Taranu2017,Bienayme2018}. Nevertheless, even the sophisticated $N$-body and DF-based approaches often do not produce systems in exact equilibrium, and need to be followed by an initial ``relaxation'' stage before running merger simulations \citep[e.g. figure~2 and section~3.2.1 of][]{Garavito2019}. On the other hand, $N$-body simulations are also subjected to numerical heating that leads to an artificial expansion of the flattened components and to a randomisation of the circular orbits that also need to be taken in consideration \citep[][hereafter L19, L21, and W22, respectively]{Ludlow2019,Ludlow2021,Wilkinson2022}.

In this paper, we present a new method for constructing $N$-body galaxy models based on DFs and tailored to observed kinematics. We apply our procedure to the NSD host FCC\,170 (NGC\,1381), for which \citeauthor{pinna2019FCC170} (\citeyear{pinna2019FCC170}; hereafter P19) constructed detailed 2D maps of stellar ages, abundances, and stellar kinematics along the line of sight (LOS). We demonstrate that these models reproduce well the observed data and are in near-perfect equilibrium, eliminating the need to perform an initial relaxation stage. We also show how numerical heating affects the flattening of both the thin disc and the NSD of our $N$-body FCC\,170 model. With the mass resolution achieved in our tests, this numerical artefact is well under control for both the kpc-scale and nuclear disc of FCC\,170, introducing an artificial 2 per cent and 20 per cent thickening of such discs, respectively, after 10~Gyr of isolated evolution. The current study is the first step toward the exploration of the fragility of thin-disc structures (kpc-scale and nuclear scale) against different galactic mergers, that will be the topic of a follow-up paper currently in preparation (hereafter Paper~II).

The paper is structured as follows: in Section~\ref{sec:observations}, we give a brief description of the FCC\,170 data obtained from the Multi Unit Spectroscopic Explorer (MUSE) and its observed kinematics. Section~\ref{sec:df_model} explains how we build the pure $N$-body models based on DFs. In Section~\ref{sec:Gizmo}, we describe the setup of the code used to evolve the $N$-body model, along with the discussion on the stability and evolution of the $N$-body galaxy in isolation. In Section~\ref{sec:conclusions}, we give our conclusions.

%--------------------------------------------------------------------
\section{Spectroscopic observations and stellar kinematics of FCC~170}\label{sec:observations}

We selected the edge-on galaxy FCC\,170 (NGC\,1381), hosted in the Fornax cluster, making use of the data obtained by the Fornax3D survey \citep{sarzi2018,iodice2019} with the MUSE integral-field unit installed at the Very Large Telescope. The MUSE datacubes were taken using the wide-field mode, providing a spatial sampling of $0.2\arcsec\times0.2\arcsec$ on a $1\arcmin\times1\arcmin$ field of view. The wavelength range of the MUSE datacubes is enclosed between 4650 and 9300~\AA, with a spectral sampling of $1.25$ \AA\ pixel$^{-1}$ and an average spectral resolution FWHM$_{\rm int}=2.8$~\AA.  The spatial scaling of the MUSE images is $0.2\arcsec \,{\rm px}^{-1}$ and we assume a distance towards our target galaxy FCC\,170 of 21.9~Mpc, in concordance to P19. The actual field of view consists of a mosaic of two pointings: a central pointing that covers the inner regions of the galaxy and an offset pointing that covers the outer disc and halo region of the galaxy. The central and offset pointings have integration times of 60 min and 90 min, respectively, due to different signal-to-noise ratio (SNR), in order to reach the same limiting surface brightness of $\mu_{\rm B}=25 \rm \,mag\,arcsec^{-2}$. The acquisition and reduction of the datacubes is extensively described in \citet{sarzi2018} and \citet{iodice2019}.

The MUSE pointings have been reduced using their own dedicated pipeline \citep{Weilbacher2012, Weilbacher2016} within the environment ESOREFLEX \citep{Freudling2013}, as described in \citet{sarzi2018} and \citet{iodice2019}. In the reduction, they took care of sky subtraction, telluric correction, and flux calibration, either relative and absolute. Generally, single pointings are aligned throughout reference stars and later combined to produce the final MUSE mosaics.

P19 and \citet{iodice2019} present 2D maps of the stellar kinematics for the edge-on lenticular galaxy FCC\,170, showing also maps of the stellar populations including metallicities and ages. The main differences between these studies are the required target SNR of the Voronoi bin (P19 impose a target $\rm SNR=40$ for FCC\,170 and a minimum $\rm SNR=1$ per spaxel, whereas \citealt{iodice2019} require a minimum $\rm SNR=3$ per spaxel) and the methodology for deriving stellar populations based on two different methods: spectral fitting in P19 and measurement of line-strength indices in \citet{iodice2019}. P19 obtained kinematics 2D maps of FCC\,170 by using the Penalized Pixel-Fitting algorithm \citep[pPXF;][]{cappellari2004,cappellari2017}, fitting the stellar spectra of the galaxy combined with a set of stellar population templates. The galaxy is initially binned in a Voronoi tessellation grid \citep{cappellari2003} accounting for a required minimum SNR in each spaxel. After binning the flux image of the galaxy, P19 apply a set of single stellar population templates from the MILES stellar library based on BaSTI isochrones \citep[described in][]{Vazdekis2015} to fit each Voronoi-binned spectra and recover the kinematics of the stars, using a standard Gauss--Hermite expansion to parametrize the stellar LOS velocity distribution \citep[LOSVD;][]{gerhard1993, vanderMarel1993}.

% -------------------------------------------------------------------------
\section{Building {\it N}-body models tailored to FCC~170 observations}\label{sec:df_model}

\subsection{Method}

We construct initial conditions for our simulations resembling the actual FCC\,170 galaxy, using the \textsc{Agama} stellar-dynamical framework \citep{vasiliev2019}. It provides a wide range of tools for various tasks in stellar dynamics, in particular, several methods for constructing multicomponent equilibrium galaxy models and computing their observational properties. We use the iterative self-consistent modelling approach described in section~6.2 of that paper. Below we briefly summarise its features and our fitting strategy.

According to the Jeans theorem, in the dynamical equilibrium, the DF of each galactic population may only depend on the integrals of motion. In this approach, we use actions $\boldsymbol J$ as the integrals of motion, relying on the St\"ackel approximation \citep{Binney2012} for mapping between the position-velocity ($\boldsymbol x, \boldsymbol v$) and action spaces, which in the current version is limited to axisymmetric systems. After choosing a suitable functional form  of each component's DF $f_{c}(\boldsymbol J)$, as described below, the construction of a self-consistent equilibrium model proceeds iteratively. We adopt a plausible initial guess for the gravitational potential $\Phi(\boldsymbol x)$, then compute the density profile generated by each component's DF,

\begin{align}
\rho_c(\boldsymbol x) \equiv \iiint \mathrm{d}^3\boldsymbol x\; f_{c}\big(\boldsymbol J(\boldsymbol x, \boldsymbol v;\;\Phi)\big),\\ \nonumber
\end{align}

\noindent and finally recompute the total potential $\Phi$ from the Poisson equation,
$\nabla^2\Phi = 4\pi G\sum_c \rho_c$, where $G$ is the gravitational constant. The whole process is then repeated several times, until the changes in the potential are negligible ($\lesssim 1$~per cent) \citep{Binney2014}.

We use different classes of DFs for disc and spheroidal galactic components, which are fully described in section~4 of \citet{vasiliev2019}. The three discs (thin, thick, and NSD) are represented by \texttt{QuasiIsothermal} models with the following free parameters: scale length, scale height, central value of the radial velocity dispersion, and scale radius of its exponential decay. The radial and vertical density profiles generated by this DF are close to exponential and sech$^2$ \citep{Kruit2011}, respectively, as it is typical for stellar discs. The bulge DF belongs to the \texttt{DoublePowerLaw} family \citep{Posti2015}, which has as many as nine free parameters: power-law indices of asymptotic behaviour at small and large radii, steepness of the transition between the two regimes, corresponding characteristic spatial scale, amount of rotation, and four dimensionless coefficients describing the radial and vertical velocity anisotropy (the latter also implicitly determines the flattening of the density profile) at small and large radii. Finally, for the dark matter (DM) halo we do not have any kinematic constraints, so we choose a simple \texttt{QuasiSpherical} DF determined by its density profile, which is taken to follow a \citet{Navarro1996} model with an adjustable scale length and a truncation radius 10 times larger (it is essentially unconstrained by stellar kinematics). The masses of all five components are also free parameters, bringing their total number to 26.

We also add a central SMBH with a mass $3 \times 10^7$~M$_{\sun}$. It is slightly larger than the value adopted by \citealt{Poci2021}, but consistent with the SMBH mass--spheroid mass relation as given by \citet{Kormendy2013}. This value is fixed throughout the fitting process, since the expected kinematic signature of such an SMBH is confined to a very small spatial region (the sphere of influence), which is not resolved by the non-adaptive-optics integral-field unit observations.

\subsection{Fitting models to observations}

\begin{figure*}
\includegraphics{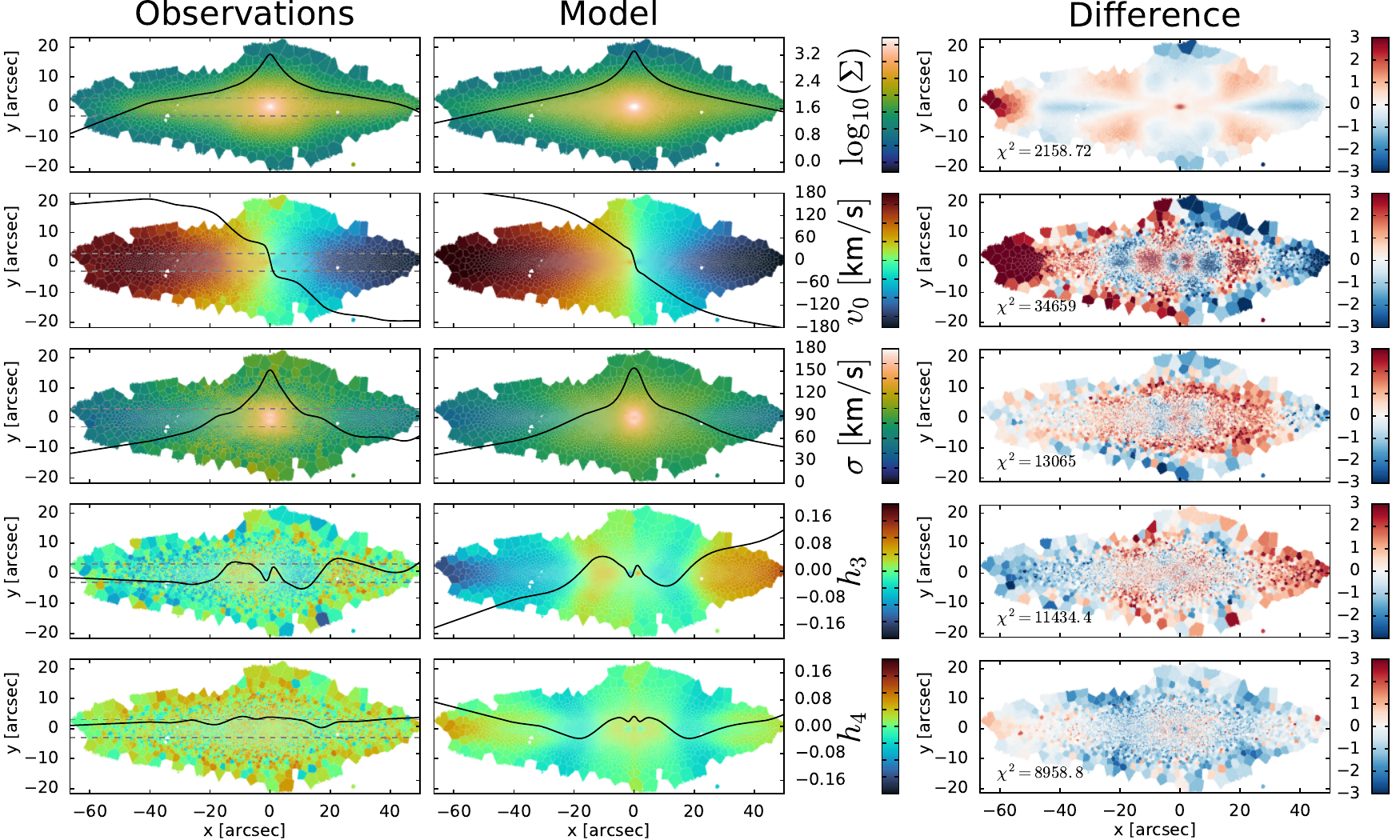}
\caption{Comparison of the observational data (left-hand column) with the DF-based model (central column) and (observed minus model) residuals normalized by the observational uncertainties (right-hand column) along with the $\chi^2$ of each residual map. From top to bottom, the rows show the maps of the stellar surface luminosity density and the four parameters describing the LOSVD in terms of Gauss--Hermite moments. The solid lines correspond to the radial profiles extracted along the galaxy major axis in the region bracketed by the horizontal dashed lines. 
}  
\label{fig:FCC170_kin_maps}
\end{figure*}

To evaluate the likelihood of a model with the given set of parameters, we need to compute the LOSVDs in each of the $\sim$8000 Voronoi bins, convert them to Gauss--Hermite moments, and then compare to the observational kinematic maps and to the surface brightness map.
The parameters of the Gauss--Hermite expansion are the centre value $v_0$ and width $\sigma$ of the base Gaussian (which are close to but not identical to the mean velocity and its dispersion, respectively), and two higher-order moments $h_3$ and $h_4$ (which are related to skewness and kurtosis and quantify the shape of the LOSVD).
Although \textsc{Agama} can compute these quantities very accurately by integrating the DF over the LOS and the two sky-plane velocity components, this is very computationally expensive. Instead, we use the following strategy: the DFs of a fiducial model $f_c^\mathrm{(fid)}$ are sampled into equal-mass $N$-body snapshots, and the positions/velocities $\{\boldsymbol x_k, \boldsymbol v_k\}$ of these particles are used to compute the kinematic maps in the Voronoi bins (essentially replacing the deterministic multidimensional integration by Monte Carlo sampling). The rejection sampling procedure itself  requires a few times larger number of DF evaluations than the number of output samples, and inevitably introduces some discreteness noise due to a finite number of particles. To mitigate both factors, in the subsequent evaluation of model likelihoods for different choices of parameters, we use the same set of sample points in the 6D phase space, but reweigh their contribution to the kinematic maps by the ratio of the DF values in the current model to those in the fiducial one, $f_c^\mathrm{(curr)}\big( \boldsymbol J(\boldsymbol x_k, \boldsymbol v_k;\;\Phi^\mathrm{(curr)}) \big) / f_c^\mathrm{(fid)}\big( \boldsymbol J(\boldsymbol x_k, \boldsymbol v_k;\;\Phi^\mathrm{(fid)}) \big)$, where the conversion from the phase space to action space also depends on the corresponding model potential. Finally, in the early stages of the parameter space exploration, we use a more approximate but $\sim$$5\times$ faster interpolation scheme for action computation, and only switch to the more accurate default (non-interpolated) approach once near the maximum-likelihood point. In the end, each model construction and likelihood evaluation on a 32-core workstation takes only $\mathcal O(10)$ seconds with interpolated actions, or about a minute in the default approach, making it possible to explore thousands of points in the parameter space.

We use a standard simplex (Nelder--Mead) method for minimising the objective function, but find it to be struggling with the high dimensionality of the parameter space and inevitable noisiness of the likelihood function stemming from various numerical effects. To reduce the chance of being trapped in a local minimum, we restarted the optimisation procedure multiple times from various initial points; the fits usually ended up in the same region, but the rate of convergence was fairly slow. Evidently, a more efficient optimisation scheme is needed to make this fitting method practical.

\placetabAGAMA

Figure~\ref{fig:FCC170_kin_maps} compares the kinematic maps of the ``best-fitting'' (in a loosely defined sense) model with observations. The surface brightness map is well reproduced, even though  the stellar density profile is not fitted independently, but comes out as an integral of the DF over velocity. The kinematic maps also qualitatively match the data, but not without apparent deviations in all four Gauss--Hermite moments. The most noticeable discrepancy occurs in the bulge region outside the NSD, up to a few arcseconds (corresponding to a few hundred pc): the model fails to reproduce a nearly flat plateau in $v_0\approx\pm 60\,\mathrm{km\,s}^{-1}$ between $1''$ and $5''$ and a subsequent rather sharp rise of $v_0$ to more than twice this value in the disc region. Likewise, the central region of high velocity dispersion extends much farther in the model than in the observations. This suggests that the parametric DF family used in our fits does not have enough flexibility to fully represent the kinematic structure of the bulge and its transition to the disc. There are also noticeable deviations in the higher Gauss--Hermite moments in the outer parts of the disc, probably due to inadequacy of the exponential decline of the velocity dispersion with radius prescribed by the disc DF. Nevertheless, the DF-based models are suitable for the main purpose of this paper -- the creation of equilibrium $N$-body models qualitatively resembling the observed galaxy and having a physically motivated multicomponent structure. A list of the main parameters we use in \textsc{Agama} to set up our FCC\,170 $N$-body model are tabulated in Table~\ref{tab:AGAMA_params}. 

We also explored an alternative approach for the construction of dynamical models tailored to observations, namely the \citet{Schwarzschild1979} orbit-superposition code \textsc{Forstand} (also included in \textsc{Agama}). A comparison between the orbit-based and DF-based models is provided in Appendix~\ref{sec:schw_model}.

%--------------------------------------------------------------------
\subsection{Comparison with the photometric decomposition}

The NSD in FCC\,170 is a thin nuclear structure at scales of a few tens of pc, as shown by \citet{ledo2010}, who catalogued a variety of NSDs with scale lengths of about $\sim$100~pc, including the one at the centre of FCC\,170. The image of the NSD, along with its magnitude, ellipticity, and $A4$ parameters are shown in the right-hand panels of figure~A3 in \citet{ledo2010}. \textcolor{blue}{Morelli et al. (in prep.)} produce an in-depth modelling of the surface brightness distribution of this NSD.
They have analysed a Wide Advanced Camera Survey image of FCC\,170 obtained from the Hubble Legacy Archive with the Scorza \& Bender surface brightness decomposition \citep{Scorza1998,Morelli2004} as implemented by \citet{Corsini2016}. This method is based on the assumption that the isophotal discyness is the result of the superposition of a spheroidal component (which is either an elliptical galaxy or a bulge) and an inclined infinitesimally thin exponential disc. The two components are assumed to have both perfectly elliptical isophotes with constant but different ellipticities.

In Figure~\ref{fig:NSD_profile}, we present the contribution $D/T$ of the NSD to the total stellar light measured in the central region along the major axis of FCC\,170 (red solid line, \textcolor{blue}{Morelli et al., in prep.}) and the total particle mass of its equivalent $N$-body model (blue dashed line). We also include the $1\sigma$ errors of the photometric decomposition with the shaded grey region. The selection of the bin size for the $N$-body model is done by adopting the characteristic scale height of the NSD (Table~\ref{tab:AGAMA_params}) as the constant bin height (0.21\arcsec), while the bin width is given by a logarithmic radius of constant step $\Delta\ln{(R/{\rm arcsec)}}=0.14$ to account for quick variations of the $D/T$ at very short scales in contrast with a smoother slope at larger scales. The $D/T$ radial profiles from the photometric decomposition and $N$-body model match each other rather well and show how good are our $N$-body simulations in reproducing such a very peculiar galaxy as FCC\,170. The small ($<5$ per cent) discrepancies in the central region of the galaxy ($r<2$\arcsec) are a consequence of the fact that the $N$-body model of the NSD is mainly based on kinematics, whereas the observational properties of the NSD result from photometry only.

\placefigNSDprofiles

\placefigFCCmodel

%--------------------------------------------------------------------
\section{Checking the stability of thin discs in isolation}\label{sec:Gizmo}

We check the stability and passive evolution of our model in isolation in order to quantify how thin-disc structures are affected by numerical heating or local dynamical instabilities. Particularly, disc galaxies could form bars \citep[e.g.][]{Abbott2017,Zana2018,Patsis2019} and spiral arms \citep[e.g.][]{diaz-garcia2019,Sellwood2019,Martinez-Medina2022} with the formation of the latter being often induced by galactic bars \citep{Garma-Oehmichen2021} or pseudo-bulges \citep{Yu2022}. Moreover, the stability test in isolation provides a good reference to assess how much the thin stellar discs are perturbed by a merging event, when we will consider the case of two interacting galaxies.

\subsection{Creating and running the {\it N}-body simulation}

After the best-fitting parameters have been found, we construct an $N$-body realisation of the model by sampling particle positions and velocities from the DF. We use particles of the same mass for all stellar components, but retain the information about the component they belong to, and use a larger particle mass for the DM halo. The SMBH is represented by a single massive particle with a very small softening length.

In order to evolve our $N$-body model in isolation, we use the code \textsc{gizmo} \citep{Hopkins2015}, which has an architecture that accounts for many different physical processes, including magneto-hydrodynamics and the black hole and supernova feedback. \textsc{gizmo} offers two different gravitational solvers -- hybrid Tree or the Tree--Particle Mesh scheme -- both offering automatic adaptivity of the gravitational resolution in structures that are collapsing or expanding. \textsc{gizmo} also permits to fix the softening lengths of each particle type independently, which facilitates to account for 2D scattering at different spatial scales.

Softening lengths have been chosen to be small enough to properly resolve the vertical structure of each component. This can normally be addressed by selecting a length at most half of the characteristic scale radius or scale height of a given component. We have chosen the softening lengths of each particle type to be: $\varepsilon_{\rm DM}=50$~pc, $\varepsilon_{\rm bulge}=\varepsilon_{\rm thin-disc}=\varepsilon_{\rm thick-disc}=10$~pc, $\varepsilon_{\rm NSD}=5$~pc and $\varepsilon_{\rm SMBH}=1$~pc. We run the simulation by using 480 processor cores spread between 15 nodes, taking about 200 hours of wall-clock time to reach a total integration time of 10~Gyr and producing 200 snapshots equally spaced in time by 50~Myr. Our main run hosts $N_{\star}=3.71\times10^{6}$ stellar particles with $M_{\star}=7.32\times10^{3}$~M$_{\sun}$ per particle and $N_{\rm DM}=1.37\times10^{7}$ DM particles with $M_{\rm DM}=2.48\times10^{5}$~M$_{\sun}$ per particle. To further test numerical heating on thin discs (see Section~\ref{sec:thin-discs}), we also run three additional $N$-body runs of FCC\,170 by decreasing the number of DM particles by 10, 5, and 2 with respect to our main model.

In Figure~\ref{fig:FCC170_model}, we show the evolution of the model by plotting the surface mass density of stellar particles at different times. The bottom panels show the face-on view of the whole stellar structure and the top panels show the edge-on view. Investigating the stellar particles, we observe that the model remains in equilibrium throughout the whole simulation without forming a bar, spiral arms, or any dynamical instability. At 10~Gyr, the disc becomes slightly thicker, increasing the vertical size and partially losing its initial flattening, as a consequence of numerical heating. We explore in more detail the thickening of both the thin galactic disc and NSD in Section~\ref{sec:thin-discs}.

Last, when evolving the model in isolation, we find that the kinematics at $t=10$~Gyr remains similar to that shown in Figure~\ref{fig:FCC170_kin_maps}, confirming that the galaxy remains stable for the whole run.

%--------------------------------------------------------------------
\subsection{Numerical heating of thin-disc components}\label{sec:thin-discs}

\placefigThinDiscHeating

\placefigNSDDiscHeating

L19, L21, and W22 demonstrated that different mass components in $N$-body simulations lead to a numerical expansion of the less massive structural component and, consequently, to a flow of kinetic energy from the more massive components to the less massive ones. This is a numerical effect in which less massive components reduce their numerical two-body relaxation time (already significantly shorter than the physical one due to the limited number of particles used) by an additional $\mu=m_{1}/m_{2}$ factor (with $m_{1} \geq m_{2}$), where $m_{1}$ and $m_{2}$ are the masses of particle type 1 and 2, respectively. As a consequence, we expect the natural size of every disc structure to be increased in both radial scale and height, as reported by L19, L21, and W22, as the mass of the stellar particles is way smaller than the mass of the DM particles. Increasing the number of particles in the other components with respect to the NSD with the aim to significantly reduce the $\mu$ fraction will alleviate (although possibly not fully solve; see L19) the issue. W22 show that, when the number of particles of the DM halo rises up to $\sim10^{7}$ for a prefixed individual stellar particle mass, numerical heating on thin-disc structures gets significantly reduced, becoming negligible in a $\sim$10~Gyr time-scale.

In Appendix~\ref{sec:isolated_NSD}, we show that numerical heating on the NSD is a direct consequence of the presence of stellar and DM particles with different mass. We replace the DM halo and the other stellar components except the NSD itself by a static analytic \citet{Hernquist1990} potential with parameters chosen to approximate their combined gravitational force. The NSD and the central SMBH are then evolved as a $N$-body system in this external potential, and in this case the NSD does not experience any significant expansion or distortion.

To investigate the effects that the other particle types have on the NSD, we re-build the FCC\,170 model using the same setup as in Table~\ref{tab:AGAMA_params} but decreasing the number of DM particles by a factor 10, 5, and 2. By running all these models in isolation, we observed how segregation depends on the differences in mass between the stellar and DM components. In Figure~\ref{fig:thin-disc_comparison}, we show the evolution of thin-disc particles through time for four runs of the FCC\,170 $N$-body model with $10^{6}$, $2\times10^{6}$, $5\times10^{6}$, and $10^{7}$ (with the latter the main model) DM particles. Figure~\ref{fig:NSD_comparison} shows the NSD particles for the same snapshots as in Figure~\ref{fig:thin-disc_comparison}. The effect of numerical heating correlates with the mass of the DM particles getting significantly reduced when $N_{\rm DM}\geq10^{7}$, in agreement with the results found by W22 (see their figures 1, 2, and 3). Indeed, the variations on the NSD are rather larger than those for the thin-disc component, as the characteristic scale length and mass of the NSD is about 10 times smaller than that of the thin disc. Consequently, as the relaxation time is proportional to both the size and number of particles, we should expect the relaxation time of the NSD to be shorter than that of the thin disc, and numerical heating would reduce this relaxation time even more. 

If these results show that numerical heating can significantly affect smaller thin-disc structures, such artificial changes may not necessarily impede our planned investigation of the impact of intermediate/minor mergers on NSDs, as long as a more precise evaluation of numerical disc heating and of its evolution in time is at hand.

\subsection{Quantifying numerical heating by mock imaging}\label{sec:photometry}

\placefigPhotometry

To quantify the time changes in shape and surface brightness of our $N$-body model of FCC\,170, we measure the profiles of ellipticity and surface mass density of the thin disc and the NSD, using the package {\textsc{photutils}} \citep{Bradley2020}. We fit a set of ellipses to edge-on projections of the thin disc and the NSD separately, excluding the other components.

In Figure~\ref{fig:photometry}, we show the photometric analysis of our $N$-body model at different times plotting the surface mass density and ellipticity of all the stellar particles (blue lines), thin-disc particles (red lines), and NSD particles (black lines) along the major axis in edge-on projection. We also indicate the limiting values of ellipticity at large radii at $t=0$ (dashed horizontal lines), in order to illustrate how this value decreases with time for each component. The dot-dashed lines show the values of surface mass density and ellipticity for the $N$-body model with $10^{6}$ DM particles. We observe that the surface mass density of the galaxy evolves very little with time for the model with $10^7$ DM particles, contrary to the lower-resolution model. All stellar components gradually expand in radius and thickness and decrease their surface mass density, as a consequence of numerical heating. To quantify the impact of heating on the disc thickness, we analyse the ellipticity of the stellar distribution. In the model with $10^7$ DM particles, the ellipticity of the entire stellar distribution decreases by just 2 per cent over the course of 10~Gyr, with the structure of the kpc-scale thin disc changing even less, by 1 per cent at most. The impact on the NSD particles is more pronounced, with the NSD eventually becoming 22 per cent thicker after 10~Gyr. On shorter time-scales, however, the impact of numerical heating is more contained, with an artificial thickening of $\sim$5, 6, and 10 per cent after 1, 2 and 5~Gyr, respectively. Figure~\ref{fig:photometry} also further demonstrates how increasing the number of DM particles is critical in controlling the impact of numerical heating on smaller central structures, as adopting only $10^{6}$ DM particles leads not only to a much thicker but also considerably more extended NSD structure.

In Paper~II, we will analyse how mergers affect the thickening of thin discs by comparing the time changes in ellipticity of the merging model with the $N$-body FCC\,170 model evolved in isolation. For instance, if at the end of a 1:4 merger event lasting $\sim$2~Gyr (counting since the first pericentre passage of the secondary galaxy), we observe a decrease in the NSD ellipticity by 25 per cent, then we may conclude that the merger event affected the NSD beyond what could be artificially produced by numerical heating.

%-----------------------------------------------------------------
\section{Conclusions}\label{sec:conclusions}

We presented a pure $N$-body model tailored to represent the edge-on lenticular galaxy FCC\,170 based on integral-field spectroscopic data. It is constructed using a new approach for building self-consistent equilibrium models specified by DFs. The DF parameters are optimized to match the observed kinematics, including not only $v$ and $\sigma$ but also the high-order moments of the LOSVD. We track the evolution of the thin disc and NSD in the subsequent $N$-body simulation. We demonstrate that the model remains in equilibrium when evolved in isolation, and its properties do not change significantly over 10~Gyr. The only exception is a moderate increase in thickness of the NSD, caused by numerical heating from the much heavier DM halo particles. We kept this artifical thickening under control by using a sufficiently high number of halo particles.

In Paper~II, we will explore the fragility of thin-disc structures in galactic encounters between our $N$-body model and a secondary spheroidal galaxy in a scenario that is consistent with the mergers observed in cosmological simulations. In this context, our $N$-body model offers an advantage compared with cosmological simulations (along with a higher resolution than various zoom-in state-of-the-art simulations), as we can control the conditions of the merger such as the initial distance between galaxies and the type of orbit.

\section*{Acknowledgements}

We thank the Reviewer, Curtis Struck, for the constructive feedback. This work was performed on the OzSTAR national facility at Swinburne University of Technology. The OzSTAR program receives funding in part from the Astronomy National Collaborative Research Infrastructure Strategy (NCRIS) allocation provided by the Australian Government. We are grateful for use of the computing resources from the Northern Ireland High Performance Computing (NI-HPC) service funded by the Engineering and Physical Sciences Research Council (EPSRC) (EP/T022175). Lorenzo Posti acknowledges support from the European Research Council (ERC) under the European Union Horizon 2020 research and innovation program (grant agreement No. 834148). Enrico Maria Corsini acknowledges support by Padua University grants DOR 2019-2022 and by Italian Ministry for Education University and Research (MIUR) grant PRIN 2017 20173ML3WW-001.

%%%%%%%%%%%%%%%%%%%%%%%%%%%%%%%%%%%%%%%%%%%%%%%%%%
\section*{Data Availability Statement}

The data underlying this article can be made available upon request. The model results can be reproduced using publicly available codes.



\bibliographystyle{mnras}
\bibliography{paper} % if your bibtex file is called example.bib

\begin{thebibliography}{}
\makeatletter
\relax
\def\mn@urlcharsother{\let\do\@makeother \do\$\do\&\do\#\do\^\do\_\do\%\do\~}
\def\mn@doi{\begingroup\mn@urlcharsother \@ifnextchar [ {\mn@doi@}
  {\mn@doi@[]}}
\def\mn@doi@[#1]#2{\def\@tempa{#1}\ifx\@tempa\@empty \href
  {http://dx.doi.org/#2} {doi:#2}\else \href {http://dx.doi.org/#2} {#1}\fi
  \endgroup}
\def\mn@eprint#1#2{\mn@eprint@#1:#2::\@nil}
\def\mn@eprint@arXiv#1{\href {http://arxiv.org/abs/#1} {{\tt arXiv:#1}}}
\def\mn@eprint@dblp#1{\href {http://dblp.uni-trier.de/rec/bibtex/#1.xml}
  {dblp:#1}}
\def\mn@eprint@#1:#2:#3:#4\@nil{\def\@tempa {#1}\def\@tempb {#2}\def\@tempc
  {#3}\ifx \@tempc \@empty \let \@tempc \@tempb \let \@tempb \@tempa \fi \ifx
  \@tempb \@empty \def\@tempb {arXiv}\fi \@ifundefined
  {mn@eprint@\@tempb}{\@tempb:\@tempc}{\expandafter \expandafter \csname
  mn@eprint@\@tempb\endcsname \expandafter{\@tempc}}}

\bibitem[\protect\citeauthoryear{{Abadi}, {Navarro}, {Steinmetz}  \&
  {Eke}}{{Abadi} et~al.}{2003}]{Abadi2003}
{Abadi} M.~G.,  {Navarro} J.~F.,  {Steinmetz} M.,   {Eke} V.~R.,  2003, \mn@doi
  [\apj] {10.1086/378316}, \href
  {https://ui.adsabs.harvard.edu/abs/2003ApJ...597...21A} {597, 21}

\bibitem[\protect\citeauthoryear{{Abbott}, {Valluri}, {Shen}  \&
  {Debattista}}{{Abbott} et~al.}{2017}]{Abbott2017}
{Abbott} C.~G.,  {Valluri} M.,  {Shen} J.,   {Debattista} V.~P.,  2017, \mn@doi
  [\mnras] {10.1093/mnras/stx1262}, \href
  {https://ui.adsabs.harvard.edu/abs/2017MNRAS.470.1526A} {470, 1526}

\bibitem[\protect\citeauthoryear{Athanassoula, Rodionov, Peschken  \&
  Lambert}{Athanassoula et~al.}{2016}]{Athanassoula2016}
Athanassoula E.,  Rodionov S.,  Peschken N.,   Lambert J.,  2016, The
  Astrophysical Journal, 821, 90

\bibitem[\protect\citeauthoryear{{Barnes} \& {Hernquist}}{{Barnes} \&
  {Hernquist}}{1992}]{barnes1992}
{Barnes} J.~E.,  {Hernquist} L.,  1992, \mn@doi [\araa]
  {10.1146/annurev.aa.30.090192.003421}, \href
  {https://ui.adsabs.harvard.edu/abs/1992ARA&A..30..705B} {30, 705}

\bibitem[\protect\citeauthoryear{{Bienaym{\'e}}, {Leca}  \&
  {Robin}}{{Bienaym{\'e}} et~al.}{2018}]{Bienayme2018}
{Bienaym{\'e}} O.,  {Leca} J.,   {Robin} A.~C.,  2018, \mn@doi [\aap]
  {10.1051/0004-6361/201833395}, \href
  {https://ui.adsabs.harvard.edu/abs/2018A&A...620A.103B} {620, A103}

\bibitem[\protect\citeauthoryear{{B{\'\i}lek}, {Fensch}, {Ebrov{\'a}},
  {Nagesh}, {Famaey}, {Duc}  \& {Kroupa}}{{B{\'\i}lek}
  et~al.}{2022}]{Bilek2022}
{B{\'\i}lek} M.,  {Fensch} J.,  {Ebrov{\'a}} I.,  {Nagesh} S.~T.,  {Famaey} B.,
   {Duc} P.-A.,   {Kroupa} P.,  2022, \mn@doi [\aap]
  {10.1051/0004-6361/202141709}, \href
  {https://ui.adsabs.harvard.edu/abs/2022A&A...660A..28B} {660, A28}

\bibitem[\protect\citeauthoryear{{Binney}}{{Binney}}{2012}]{Binney2012}
{Binney} J.,  2012, \mn@doi [\mnras] {10.1111/j.1365-2966.2012.21757.x}, \href
  {https://ui.adsabs.harvard.edu/abs/2012MNRAS.426.1324B} {426, 1324}

\bibitem[\protect\citeauthoryear{{Binney}}{{Binney}}{2014}]{Binney2014}
{Binney} J.,  2014, \mn@doi [\mnras] {10.1093/mnras/stu297}, \href
  {https://ui.adsabs.harvard.edu/abs/2014MNRAS.440..787B} {440, 787}

\bibitem[\protect\citeauthoryear{{Bradley} et~al.,}{{Bradley}
  et~al.}{2020}]{Bradley2020}
{Bradley} L.,  et~al., 2020, {astropy/photutils: 1.0.0}, Zenodo,
  \mn@doi{10.5281/zenodo.4044744}

\bibitem[\protect\citeauthoryear{{Capelo}, {Volonteri}, {Dotti}, {Bellovary},
  {Mayer}  \& {Governato}}{{Capelo} et~al.}{2015}]{capelo2015}
{Capelo} P.~R.,  {Volonteri} M.,  {Dotti} M.,  {Bellovary} J.~M.,  {Mayer} L.,
   {Governato} F.,  2015, \mn@doi [\mnras] {10.1093/mnras/stu2500}, \href
  {https://ui.adsabs.harvard.edu/abs/2015MNRAS.447.2123C} {447, 2123}

\bibitem[\protect\citeauthoryear{{Cappellari}}{{Cappellari}}{2008}]{Cappellari2008}
{Cappellari} M.,  2008, \mn@doi [\mnras] {10.1111/j.1365-2966.2008.13754.x},
  \href {https://ui.adsabs.harvard.edu/abs/2008MNRAS.390...71C} {390, 71}

\bibitem[\protect\citeauthoryear{{Cappellari}}{{Cappellari}}{2017}]{cappellari2017}
{Cappellari} M.,  2017, \mn@doi [\mnras] {10.1093/mnras/stw3020}, \href
  {https://ui.adsabs.harvard.edu/abs/2017MNRAS.466..798C} {466, 798}

\bibitem[\protect\citeauthoryear{{Cappellari} \& {Copin}}{{Cappellari} \&
  {Copin}}{2003}]{cappellari2003}
{Cappellari} M.,  {Copin} Y.,  2003, \mn@doi [\mnras]
  {10.1046/j.1365-8711.2003.06541.x}, \href
  {https://ui.adsabs.harvard.edu/abs/2003MNRAS.342..345C} {342, 345}

\bibitem[\protect\citeauthoryear{{Cappellari} \& {Emsellem}}{{Cappellari} \&
  {Emsellem}}{2004}]{cappellari2004}
{Cappellari} M.,  {Emsellem} E.,  2004, \mn@doi [\pasp] {10.1086/381875}, \href
  {https://ui.adsabs.harvard.edu/abs/2004PASP..116..138C} {116, 138}

\bibitem[\protect\citeauthoryear{{Choi}, {Park}  \& {Vogeley}}{{Choi}
  et~al.}{2007}]{choi2007}
{Choi} Y.-Y.,  {Park} C.,   {Vogeley} M.~S.,  2007, \mn@doi [\apj]
  {10.1086/511060}, \href
  {https://ui.adsabs.harvard.edu/abs/2007ApJ...658..884C} {658, 884}

\bibitem[\protect\citeauthoryear{{Corsini}}{{Corsini}}{2014}]{Corsini2014}
{Corsini} E.~M.,  2014, in {Iodice} E.,  {Corsini} E.~M.,  eds,  Astronomical
  Society of the Pacific Conference Series Vol. 486, Multi-Spin Galaxies. p.~51
  (\mn@eprint {arXiv} {1403.1263})

\bibitem[\protect\citeauthoryear{{Corsini}, {Morelli}, {Pastorello}, {Dalla
  Bont{\`a}}, {Pizzella}  \& {Portaluri}}{{Corsini} et~al.}{2016}]{Corsini2016}
{Corsini} E.~M.,  {Morelli} L.,  {Pastorello} N.,  {Dalla Bont{\`a}} E.,
  {Pizzella} A.,   {Portaluri} E.,  2016, \mn@doi [\mnras]
  {10.1093/mnras/stv2864}, \href
  {https://ui.adsabs.harvard.edu/abs/2016MNRAS.457.1198C} {457, 1198}

\bibitem[\protect\citeauthoryear{{Cretton}, {de Zeeuw}, {van der Marel}  \&
  {Rix}}{{Cretton} et~al.}{1999}]{Cretton1999}
{Cretton} N.,  {de Zeeuw} P.~T.,  {van der Marel} R.~P.,   {Rix} H.-W.,  1999,
  \mn@doi [\apjs] {10.1086/313264}, \href
  {https://ui.adsabs.harvard.edu/abs/1999ApJS..124..383C} {124, 383}

\bibitem[\protect\citeauthoryear{\DE{Lorenzi}{De}{de}~Lorenzi, {Debattista},
  {Gerhard}  \& {Sambhus}}{\DE{Lorenzi}{De}{de}~Lorenzi
  et~al.}{2007}]{deLorenzi2007}
\DE{Lorenzi}{De}{de}~Lorenzi F.,  {Debattista} V.~P.,  {Gerhard} O.,
  {Sambhus} N.,  2007, \mn@doi [\mnras] {10.1111/j.1365-2966.2007.11434.x},
  \href {https://ui.adsabs.harvard.edu/abs/2007MNRAS.376...71D} {376, 71}

\bibitem[\protect\citeauthoryear{{Davison} et~al.,}{{Davison}
  et~al.}{2021}]{Davison2021}
{Davison} T.~A.,  et~al., 2021, \mn@doi [\mnras] {10.1093/mnras/stab162}, \href
  {https://ui.adsabs.harvard.edu/abs/2021MNRAS.502.2296D} {502, 2296}

\bibitem[\protect\citeauthoryear{{Debattista} \& {Sellwood}}{{Debattista} \&
  {Sellwood}}{2000}]{Debattista2000}
{Debattista} V.~P.,  {Sellwood} J.~A.,  2000, \mn@doi [\apj] {10.1086/317148},
  \href {https://ui.adsabs.harvard.edu/abs/2000ApJ...543..704D} {543, 704}

\bibitem[\protect\citeauthoryear{{Deeley} et~al.,}{{Deeley}
  et~al.}{2017}]{Deeley2017}
{Deeley} S.,  et~al., 2017, \mn@doi [\mnras] {10.1093/mnras/stx441}, \href
  {https://ui.adsabs.harvard.edu/abs/2017MNRAS.467.3934D} {467, 3934}

\bibitem[\protect\citeauthoryear{{Dejonghe} \& {Merritt}}{{Dejonghe} \&
  {Merritt}}{1992}]{Dejonghe1992}
{Dejonghe} H.,  {Merritt} D.,  1992, \mn@doi [\apj] {10.1086/171368}, \href
  {https://ui.adsabs.harvard.edu/abs/1992ApJ...391..531D} {391, 531}

\bibitem[\protect\citeauthoryear{{D{\'\i}az-Garc{\'\i}a}, {Salo}, {Knapen}  \&
  {Herrera-Endoqui}}{{D{\'\i}az-Garc{\'\i}a} et~al.}{2019}]{diaz-garcia2019}
{D{\'\i}az-Garc{\'\i}a} S.,  {Salo} H.,  {Knapen} J.~H.,   {Herrera-Endoqui}
  M.,  2019, \mn@doi [\aap] {10.1051/0004-6361/201936000}, \href
  {https://ui.adsabs.harvard.edu/abs/2019A&A...631A..94D} {631, A94}

\bibitem[\protect\citeauthoryear{{Freudling}, {Romaniello}, {Bramich},
  {Ballester}, {Forchi}, {Garc{\'\i}a-Dabl{\'o}}, {Moehler}  \&
  {Neeser}}{{Freudling} et~al.}{2013}]{Freudling2013}
{Freudling} W.,  {Romaniello} M.,  {Bramich} D.~M.,  {Ballester} P.,  {Forchi}
  V.,  {Garc{\'\i}a-Dabl{\'o}} C.~E.,  {Moehler} S.,   {Neeser} M.~J.,  2013,
  \mn@doi [\aap] {10.1051/0004-6361/201322494}, \href
  {https://ui.adsabs.harvard.edu/abs/2013A&A...559A..96F} {559, A96}

\bibitem[\protect\citeauthoryear{{Gal{\'a}n-de Anta} et~al.,}{{Gal{\'a}n-de
  Anta} et~al.}{2022}]{Galan2022}
{Gal{\'a}n-de Anta} P.~M.,  et~al., 2022, \mn@doi [\mnras]
  {10.1093/mnras/stac3061}, \href
  {https://ui.adsabs.harvard.edu/abs/2022MNRAS.517.5992G} {517, 5992}

\bibitem[\protect\citeauthoryear{{Garavito-Camargo}, {Besla}, {Laporte},
  {Johnston}, {G{\'o}mez}  \& {Watkins}}{{Garavito-Camargo}
  et~al.}{2019}]{Garavito2019}
{Garavito-Camargo} N.,  {Besla} G.,  {Laporte} C. F.~P.,  {Johnston} K.~V.,
  {G{\'o}mez} F.~A.,   {Watkins} L.~L.,  2019, \mn@doi [\apj]
  {10.3847/1538-4357/ab32eb}, \href
  {https://ui.adsabs.harvard.edu/abs/2019ApJ...884...51G} {884, 51}

\bibitem[\protect\citeauthoryear{{Garma-Oehmichen}, {Martinez-Medina},
  {Hern{\'a}ndez-Toledo}  \& {Puerari}}{{Garma-Oehmichen}
  et~al.}{2021}]{Garma-Oehmichen2021}
{Garma-Oehmichen} L.,  {Martinez-Medina} L.,  {Hern{\'a}ndez-Toledo} H.,
  {Puerari} I.,  2021, \mn@doi [\mnras] {10.1093/mnras/stab333}, \href
  {https://ui.adsabs.harvard.edu/abs/2021MNRAS.502.4708G} {502, 4708}

\bibitem[\protect\citeauthoryear{{Gebhardt} et~al.,}{{Gebhardt}
  et~al.}{2003}]{Gebhardt2003}
{Gebhardt} K.,  et~al., 2003, \mn@doi [\apj] {10.1086/345081}, \href
  {https://ui.adsabs.harvard.edu/abs/2003ApJ...583...92G} {583, 92}

\bibitem[\protect\citeauthoryear{{Genel} et~al.,}{{Genel}
  et~al.}{2014}]{Genel2014}
{Genel} S.,  et~al., 2014, \mn@doi [\mnras] {10.1093/mnras/stu1654}, \href
  {https://ui.adsabs.harvard.edu/abs/2014MNRAS.445..175G} {445, 175}

\bibitem[\protect\citeauthoryear{{Gerhard}}{{Gerhard}}{1993}]{gerhard1993}
{Gerhard} O.~E.,  1993, \mn@doi [\mnras] {10.1093/mnras/265.1.213}, \href
  {https://ui.adsabs.harvard.edu/abs/1993MNRAS.265..213G} {265, 213}

\bibitem[\protect\citeauthoryear{{Governato}, {Willman}, {Mayer}, {Brooks},
  {Stinson}, {Valenzuela}, {Wadsley}  \& {Quinn}}{{Governato}
  et~al.}{2007}]{Governato2007}
{Governato} F.,  {Willman} B.,  {Mayer} L.,  {Brooks} A.,  {Stinson} G.,
  {Valenzuela} O.,  {Wadsley} J.,   {Quinn} T.,  2007, \mn@doi [\mnras]
  {10.1111/j.1365-2966.2006.11266.x}, \href
  {https://ui.adsabs.harvard.edu/abs/2007MNRAS.374.1479G} {374, 1479}

\bibitem[\protect\citeauthoryear{{Hammer}, {Flores}, {Yang}, {Athanassoula},
  {Puech}, {Rodrigues}  \& {Peirani}}{{Hammer} et~al.}{2009}]{Hammer2009}
{Hammer} F.,  {Flores} H.,  {Yang} Y.~B.,  {Athanassoula} E.,  {Puech} M.,
  {Rodrigues} M.,   {Peirani} S.,  2009, \mn@doi [\aap]
  {10.1051/0004-6361:200810488}, \href
  {https://ui.adsabs.harvard.edu/abs/2009A&A...496..381H} {496, 381}

\bibitem[\protect\citeauthoryear{{Hernquist}}{{Hernquist}}{1990}]{Hernquist1990}
{Hernquist} L.,  1990, \mn@doi [\apj] {10.1086/168845}, \href
  {https://ui.adsabs.harvard.edu/abs/1990ApJ...356..359H} {356, 359}

\bibitem[\protect\citeauthoryear{{Hernquist} \& {Quinn}}{{Hernquist} \&
  {Quinn}}{1988}]{Hernquist1988}
{Hernquist} L.,  {Quinn} P.~J.,  1988, \mn@doi [\apj] {10.1086/166592}, \href
  {https://ui.adsabs.harvard.edu/abs/1988ApJ...331..682H} {331, 682}

\bibitem[\protect\citeauthoryear{{Hopkins}}{{Hopkins}}{2015}]{Hopkins2015}
{Hopkins} P.~F.,  2015, \mn@doi [\mnras] {10.1093/mnras/stv195}, \href
  {https://ui.adsabs.harvard.edu/abs/2015MNRAS.450...53H} {450, 53}

\bibitem[\protect\citeauthoryear{{Ilbert} et~al.,}{{Ilbert}
  et~al.}{2006}]{ilbert2006}
{Ilbert} O.,  et~al., 2006, \mn@doi [\aap] {10.1051/0004-6361:20053632}, \href
  {https://ui.adsabs.harvard.edu/abs/2006A&A...453..809I} {453, 809}

\bibitem[\protect\citeauthoryear{{Iodice} et~al.,}{{Iodice}
  et~al.}{2019}]{iodice2019}
{Iodice} E.,  et~al., 2019, \mn@doi [\aap] {10.1051/0004-6361/201935721}, \href
  {https://ui.adsabs.harvard.edu/abs/2019A&A...627A.136I} {627, A136}

\bibitem[\protect\citeauthoryear{{Jaffe}, {Ford}, {O'Connell}, {van den Bosch}
  \& {Ferrarese}}{{Jaffe} et~al.}{1994}]{Jaffe1994}
{Jaffe} W.,  {Ford} H.~C.,  {O'Connell} R.~W.,  {van den Bosch} F.~C.,
  {Ferrarese} L.,  1994, \mn@doi [\aj] {10.1086/117178}, \href
  {https://ui.adsabs.harvard.edu/abs/1994AJ....108.1567J} {108, 1567}

\bibitem[\protect\citeauthoryear{{Joshi}, {Pillepich}, {Nelson}, {Marinacci},
  {Springel}, {Rodriguez-Gomez}, {Vogelsberger}  \& {Hernquist}}{{Joshi}
  et~al.}{2020}]{Joshi2020}
{Joshi} G.~D.,  {Pillepich} A.,  {Nelson} D.,  {Marinacci} F.,  {Springel} V.,
  {Rodriguez-Gomez} V.,  {Vogelsberger} M.,   {Hernquist} L.,  2020, \mn@doi
  [\mnras] {10.1093/mnras/staa1668}, \href
  {https://ui.adsabs.harvard.edu/abs/2020MNRAS.496.2673J} {496, 2673}

\bibitem[\protect\citeauthoryear{{Kazantzidis}, {Magorrian}  \&
  {Moore}}{{Kazantzidis} et~al.}{2004}]{Kazantzidis2004}
{Kazantzidis} S.,  {Magorrian} J.,   {Moore} B.,  2004, \mn@doi [\apj]
  {10.1086/380192}, \href
  {https://ui.adsabs.harvard.edu/abs/2004ApJ...601...37K} {601, 37}

\bibitem[\protect\citeauthoryear{{Kormendy} \& {Ho}}{{Kormendy} \&
  {Ho}}{2013}]{Kormendy2013}
{Kormendy} J.,  {Ho} L.~C.,  2013, \mn@doi [ARA\&A]
  {10.1146/annurev-astro-082708-101811}, \href
  {https://ui.adsabs.harvard.edu/abs/2013ARA&A..51..511K} {51, 511}

\bibitem[\protect\citeauthoryear{{Kuijken} \& {Dubinski}}{{Kuijken} \&
  {Dubinski}}{1995}]{Kuijken1995}
{Kuijken} K.,  {Dubinski} J.,  1995, \mn@doi [\mnras]
  {10.1093/mnras/277.4.1341}, \href
  {https://ui.adsabs.harvard.edu/abs/1995MNRAS.277.1341K} {277, 1341}

\bibitem[\protect\citeauthoryear{{Ledo}, {Sarzi}, {Dotti}, {Khochfar}  \&
  {Morelli}}{{Ledo} et~al.}{2010}]{ledo2010}
{Ledo} H.~R.,  {Sarzi} M.,  {Dotti} M.,  {Khochfar} S.,   {Morelli} L.,  2010,
  \mn@doi [\mnras] {10.1111/j.1365-2966.2010.16990.x}, \href
  {https://ui.adsabs.harvard.edu/abs/2010MNRAS.407..969L} {407, 969}

\bibitem[\protect\citeauthoryear{{{\L}okas} \& {Mamon}}{{{\L}okas} \&
  {Mamon}}{2003}]{Lokas2003}
{{\L}okas} E.~L.,  {Mamon} G.~A.,  2003, \mn@doi [\mnras]
  {10.1046/j.1365-8711.2003.06684.x}, \href
  {https://ui.adsabs.harvard.edu/abs/2003MNRAS.343..401L} {343, 401}

\bibitem[\protect\citeauthoryear{{Lotz}, {Jonsson}, {Cox}  \& {Primack}}{{Lotz}
  et~al.}{2010}]{Lotz2010}
{Lotz} J.~M.,  {Jonsson} P.,  {Cox} T.~J.,   {Primack} J.~R.,  2010, \mn@doi
  [\mnras] {10.1111/j.1365-2966.2010.16268.x}, \href
  {https://ui.adsabs.harvard.edu/abs/2010MNRAS.404..575L} {404, 575}

\bibitem[\protect\citeauthoryear{{Ludlow}, {Schaye}, {Schaller}  \&
  {Richings}}{{Ludlow} et~al.}{2019}]{Ludlow2019}
{Ludlow} A.~D.,  {Schaye} J.,  {Schaller} M.,   {Richings} J.,  2019, \mn@doi
  [\mnras] {10.1093/mnrasl/slz110}, \href
  {https://ui.adsabs.harvard.edu/abs/2019MNRAS.488L.123L} {488, L123}

\bibitem[\protect\citeauthoryear{{Ludlow}, {Fall}, {Schaye}  \&
  {Obreschkow}}{{Ludlow} et~al.}{2021}]{Ludlow2021}
{Ludlow} A.~D.,  {Fall} S.~M.,  {Schaye} J.,   {Obreschkow} D.,  2021, \mn@doi
  [\mnras] {10.1093/mnras/stab2770}, \href
  {https://ui.adsabs.harvard.edu/abs/2021MNRAS.508.5114L} {508, 5114}

\bibitem[\protect\citeauthoryear{{Mamon}, {Biviano}  \& {Bou{\'e}}}{{Mamon}
  et~al.}{2013}]{Mamon2013}
{Mamon} G.~A.,  {Biviano} A.,   {Bou{\'e}} G.,  2013, \mn@doi [\mnras]
  {10.1093/mnras/sts565}, \href
  {https://ui.adsabs.harvard.edu/abs/2013MNRAS.429.3079M} {429, 3079}

\bibitem[\protect\citeauthoryear{{Martinez-Medina}, {P{\'e}rez-Villegas}  \&
  {Peimbert}}{{Martinez-Medina} et~al.}{2022}]{Martinez-Medina2022}
{Martinez-Medina} L.,  {P{\'e}rez-Villegas} A.,   {Peimbert} A.,  2022, \mn@doi
  [\mnras] {10.1093/mnras/stac642}, \href
  {https://ui.adsabs.harvard.edu/abs/2022MNRAS.512.1574M} {512, 1574}

\bibitem[\protect\citeauthoryear{{Mazzilli Ciraulo}, {Melchior}, {Maschmann},
  {Katkov}, {Halle}, {Combes}, {Gelfand}  \& {Al Yazeedi}}{{Mazzilli Ciraulo}
  et~al.}{2021}]{Mazzilli2021}
{Mazzilli Ciraulo} B.,  {Melchior} A.-L.,  {Maschmann} D.,  {Katkov} I.~Y.,
  {Halle} A.,  {Combes} F.,  {Gelfand} J.~D.,   {Al Yazeedi} A.,  2021, \mn@doi
  [\aap] {10.1051/0004-6361/202141319}, \href
  {https://ui.adsabs.harvard.edu/abs/2021A&A...653A..47M} {653, A47}

\bibitem[\protect\citeauthoryear{{Morelli} et~al.,}{{Morelli}
  et~al.}{2004}]{Morelli2004}
{Morelli} L.,  et~al., 2004, \mn@doi [\mnras]
  {10.1111/j.1365-2966.2004.08236.x}, \href
  {https://ui.adsabs.harvard.edu/abs/2004MNRAS.354..753M} {354, 753}

\bibitem[\protect\citeauthoryear{{Moster}, {Macci{\`o}}, {Somerville},
  {Johansson}  \& {Naab}}{{Moster} et~al.}{2010}]{Moster2010}
{Moster} B.~P.,  {Macci{\`o}} A.~V.,  {Somerville} R.~S.,  {Johansson} P.~H.,
  {Naab} T.,  2010, \mn@doi [\mnras] {10.1111/j.1365-2966.2009.16190.x}, \href
  {https://ui.adsabs.harvard.edu/abs/2010MNRAS.403.1009M} {403, 1009}

\bibitem[\protect\citeauthoryear{{Naab}, {Jesseit}  \& {Burkert}}{{Naab}
  et~al.}{2006}]{Naab2006}
{Naab} T.,  {Jesseit} R.,   {Burkert} A.,  2006, \mn@doi [\mnras]
  {10.1111/j.1365-2966.2006.10902.x}, \href
  {https://ui.adsabs.harvard.edu/abs/2006MNRAS.372..839N} {372, 839}

\bibitem[\protect\citeauthoryear{{Navarro}, {Frenk}  \& {White}}{{Navarro}
  et~al.}{1996}]{Navarro1996}
{Navarro} J.~F.,  {Frenk} C.~S.,   {White} S. D.~M.,  1996, \mn@doi [\apj]
  {10.1086/177173}, \href
  {https://ui.adsabs.harvard.edu/abs/1996ApJ...462..563N} {462, 563}

\bibitem[\protect\citeauthoryear{{Onaka}, {Nakamura}, {Sakon}, {Wu}, {Ohsawa},
  {Kaneda}, {Lebouteiller}  \& {Roellig}}{{Onaka} et~al.}{2018}]{Onaka2018}
{Onaka} T.,  {Nakamura} T.,  {Sakon} I.,  {Wu} R.,  {Ohsawa} R.,  {Kaneda} H.,
  {Lebouteiller} V.,   {Roellig} T.~L.,  2018, \mn@doi [\apj]
  {10.3847/1538-4357/aaa004}, \href
  {https://ui.adsabs.harvard.edu/abs/2018ApJ...853...31O} {853, 31}

\bibitem[\protect\citeauthoryear{{Park}, {Choi}, {Vogeley}, {Gott}, {Blanton}
  \& {SDSS Collaboration}}{{Park} et~al.}{2007}]{park2007}
{Park} C.,  {Choi} Y.-Y.,  {Vogeley} M.~S.,  {Gott} J.~Richard I.,  {Blanton}
  M.~R.,   {SDSS Collaboration} 2007, \mn@doi [\apj] {10.1086/511059}, \href
  {https://ui.adsabs.harvard.edu/abs/2007ApJ...658..898P} {658, 898}

\bibitem[\protect\citeauthoryear{{Patsis} \& {Athanassoula}}{{Patsis} \&
  {Athanassoula}}{2019}]{Patsis2019}
{Patsis} P.~A.,  {Athanassoula} E.,  2019, \mn@doi [\mnras]
  {10.1093/mnras/stz2588}, \href
  {https://ui.adsabs.harvard.edu/abs/2019MNRAS.490.2740P} {490, 2740}

\bibitem[\protect\citeauthoryear{{Peschken}, {{\L}okas}  \&
  {Athanassoula}}{{Peschken} et~al.}{2020}]{Peschken2020}
{Peschken} N.,  {{\L}okas} E.~L.,   {Athanassoula} E.,  2020, \mn@doi [\mnras]
  {10.1093/mnras/staa299}, \href
  {https://ui.adsabs.harvard.edu/abs/2020MNRAS.493.1375P} {493, 1375}

\bibitem[\protect\citeauthoryear{{Piffl}, {Penoyre}  \& {Binney}}{{Piffl}
  et~al.}{2015}]{Piffl2015}
{Piffl} T.,  {Penoyre} Z.,   {Binney} J.,  2015, \mn@doi [\mnras]
  {10.1093/mnras/stv938}, \href
  {https://ui.adsabs.harvard.edu/abs/2015MNRAS.451..639P} {451, 639}

\bibitem[\protect\citeauthoryear{{Pinna} et~al.,}{{Pinna}
  et~al.}{2019}]{pinna2019FCC170}
{Pinna} F.,  et~al., 2019, \mn@doi [\aap] {10.1051/0004-6361/201833193}, \href
  {https://ui.adsabs.harvard.edu/abs/2019A&A...623A..19P} {623, A19}

\bibitem[\protect\citeauthoryear{{Pizzella}, {Corsini}, {Morelli}, {Sarzi},
  {Scarlata}, {Stiavelli}  \& {Bertola}}{{Pizzella}
  et~al.}{2002}]{Pizzella2002}
{Pizzella} A.,  {Corsini} E.~M.,  {Morelli} L.,  {Sarzi} M.,  {Scarlata} C.,
  {Stiavelli} M.,   {Bertola} F.,  2002, \mn@doi [\apj] {10.1086/340486}, \href
  {https://ui.adsabs.harvard.edu/abs/2002ApJ...573..131P} {573, 131}

\bibitem[\protect\citeauthoryear{{Poci} et~al.,}{{Poci}
  et~al.}{2021}]{Poci2021}
{Poci} A.,  et~al., 2021, \mn@doi [\aap] {10.1051/0004-6361/202039644}, \href
  {https://ui.adsabs.harvard.edu/abs/2021A&A...647A.145P} {647, A145}

\bibitem[\protect\citeauthoryear{{Pop}, {Pillepich}, {Amorisco}  \&
  {Hernquist}}{{Pop} et~al.}{2018}]{Pop2018}
{Pop} A.-R.,  {Pillepich} A.,  {Amorisco} N.~C.,   {Hernquist} L.,  2018,
  \mn@doi [\mnras] {10.1093/mnras/sty1932}, \href
  {https://ui.adsabs.harvard.edu/abs/2018MNRAS.480.1715P} {480, 1715}

\bibitem[\protect\citeauthoryear{{Posti}, {Binney}, {Nipoti}  \&
  {Ciotti}}{{Posti} et~al.}{2015}]{Posti2015}
{Posti} L.,  {Binney} J.,  {Nipoti} C.,   {Ciotti} L.,  2015, \mn@doi [\mnras]
  {10.1093/mnras/stu2608}, \href
  {https://ui.adsabs.harvard.edu/abs/2015MNRAS.447.3060P} {447, 3060}

\bibitem[\protect\citeauthoryear{{Purcell}, {Kazantzidis}  \&
  {Bullock}}{{Purcell} et~al.}{2009}]{Purcell2009}
{Purcell} C.~W.,  {Kazantzidis} S.,   {Bullock} J.~S.,  2009, \mn@doi [\apjl]
  {10.1088/0004-637X/694/2/L98}, \href
  {https://ui.adsabs.harvard.edu/abs/2009ApJ...694L..98P} {694, L98}

\bibitem[\protect\citeauthoryear{{Read} \& {Steger}}{{Read} \&
  {Steger}}{2017}]{Read2017}
{Read} J.~I.,  {Steger} P.,  2017, \mn@doi [\mnras] {10.1093/mnras/stx1798},
  \href {https://ui.adsabs.harvard.edu/abs/2017MNRAS.471.4541R} {471, 4541}

\bibitem[\protect\citeauthoryear{{Richardson} \& {Fairbairn}}{{Richardson} \&
  {Fairbairn}}{2013}]{Richardson2013}
{Richardson} T.,  {Fairbairn} M.,  2013, \mn@doi [\mnras]
  {10.1093/mnras/stt686}, \href
  {https://ui.adsabs.harvard.edu/abs/2013MNRAS.432.3361R} {432, 3361}

\bibitem[\protect\citeauthoryear{{Robertson}, {Cox}, {Hernquist}, {Franx},
  {Hopkins}, {Martini}  \& {Springel}}{{Robertson}
  et~al.}{2006}]{Robertson2006b}
{Robertson} B.,  {Cox} T.~J.,  {Hernquist} L.,  {Franx} M.,  {Hopkins} P.~F.,
  {Martini} P.,   {Springel} V.,  2006, \mn@doi [\apj] {10.1086/500360}, \href
  {https://ui.adsabs.harvard.edu/abs/2006ApJ...641...21R} {641, 21}

\bibitem[\protect\citeauthoryear{{Romanowsky}, {Strader}, {Brodie}, {Mihos},
  {Spitler}, {Forbes}, {Foster}  \& {Arnold}}{{Romanowsky}
  et~al.}{2012}]{Romanowsky2012}
{Romanowsky} A.~J.,  {Strader} J.,  {Brodie} J.~P.,  {Mihos} J.~C.,  {Spitler}
  L.~R.,  {Forbes} D.~A.,  {Foster} C.,   {Arnold} J.~A.,  2012, \mn@doi [\apj]
  {10.1088/0004-637X/748/1/29}, \href
  {https://ui.adsabs.harvard.edu/abs/2012ApJ...748...29R} {748, 29}

\bibitem[\protect\citeauthoryear{{Sarzi}, {Ledo}  \& {Dotti}}{{Sarzi}
  et~al.}{2015}]{sarzi2015}
{Sarzi} M.,  {Ledo} H.~R.,   {Dotti} M.,  2015, \mn@doi [\mnras]
  {10.1093/mnras/stv1497}, \href
  {https://ui.adsabs.harvard.edu/abs/2015MNRAS.453.1070S} {453, 1070}

\bibitem[\protect\citeauthoryear{{Sarzi} et~al.,}{{Sarzi}
  et~al.}{2016}]{Sarzi2016}
{Sarzi} M.,  et~al., 2016, \mn@doi [\mnras] {10.1093/mnras/stw099}, \href
  {https://ui.adsabs.harvard.edu/abs/2016MNRAS.457.1804S} {457, 1804}

\bibitem[\protect\citeauthoryear{{Sarzi} et~al.,}{{Sarzi}
  et~al.}{2018}]{sarzi2018}
{Sarzi} M.,  et~al., 2018, \mn@doi [\aap] {10.1051/0004-6361/201833137}, \href
  {http://adsabs.harvard.edu/abs/2018A%26A...616A.121S} {616, A121}

\bibitem[\protect\citeauthoryear{{Schwarzschild}}{{Schwarzschild}}{1979}]{Schwarzschild1979}
{Schwarzschild} M.,  1979, \mn@doi [\apj] {10.1086/157282}, \href
  {https://ui.adsabs.harvard.edu/abs/1979ApJ...232..236S} {232, 236}

\bibitem[\protect\citeauthoryear{{Scorza} \& {van den Bosch}}{{Scorza} \& {van
  den Bosch}}{1998}]{Scorza1998}
{Scorza} C.,  {van den Bosch} F.~C.,  1998, \mn@doi [\mnras]
  {10.1046/j.1365-8711.1998.01922.x}, \href
  {https://ui.adsabs.harvard.edu/abs/1998MNRAS.300..469S} {300, 469}

\bibitem[\protect\citeauthoryear{{Sellwood} \& {Carlberg}}{{Sellwood} \&
  {Carlberg}}{2019}]{Sellwood2019}
{Sellwood} J.~A.,  {Carlberg} R.~G.,  2019, \mn@doi [\mnras]
  {10.1093/mnras/stz2132}, \href
  {https://ui.adsabs.harvard.edu/abs/2019MNRAS.489..116S} {489, 116}

\bibitem[\protect\citeauthoryear{{Sijacki}, {Vogelsberger}, {Genel},
  {Springel}, {Torrey}, {Snyder}, {Nelson}  \& {Hernquist}}{{Sijacki}
  et~al.}{2015}]{Sijacki2015}
{Sijacki} D.,  {Vogelsberger} M.,  {Genel} S.,  {Springel} V.,  {Torrey} P.,
  {Snyder} G.~F.,  {Nelson} D.,   {Hernquist} L.,  2015, \mn@doi [\mnras]
  {10.1093/mnras/stv1340}, \href
  {https://ui.adsabs.harvard.edu/abs/2015MNRAS.452..575S} {452, 575}

\bibitem[\protect\citeauthoryear{{Sparre} \& {Springel}}{{Sparre} \&
  {Springel}}{2017}]{Sparre2017}
{Sparre} M.,  {Springel} V.,  2017, \mn@doi [\mnras] {10.1093/mnras/stx1516},
  \href {https://ui.adsabs.harvard.edu/abs/2017MNRAS.470.3946S} {470, 3946}

\bibitem[\protect\citeauthoryear{{Springel}}{{Springel}}{2005}]{Springel2005}
{Springel} V.,  2005, \mn@doi [\mnras] {10.1111/j.1365-2966.2005.09655.x},
  \href {https://ui.adsabs.harvard.edu/abs/2005MNRAS.364.1105S} {364, 1105}

\bibitem[\protect\citeauthoryear{{Syer} \& {Tremaine}}{{Syer} \&
  {Tremaine}}{1996}]{Syer1996}
{Syer} D.,  {Tremaine} S.,  1996, \mn@doi [\mnras] {10.1093/mnras/282.1.223},
  \href {https://ui.adsabs.harvard.edu/abs/1996MNRAS.282..223S} {282, 223}

\bibitem[\protect\citeauthoryear{{Taranu}, {Dubinski}  \& {Yee}}{{Taranu}
  et~al.}{2013}]{Taranu2013}
{Taranu} D.~S.,  {Dubinski} J.,   {Yee} H.~K.~C.,  2013, \mn@doi [\apj]
  {10.1088/0004-637X/778/1/61}, \href
  {https://ui.adsabs.harvard.edu/abs/2013ApJ...778...61T} {778, 61}

\bibitem[\protect\citeauthoryear{{Taranu} et~al.,}{{Taranu}
  et~al.}{2017}]{Taranu2017}
{Taranu} D.~S.,  et~al., 2017, \mn@doi [\apj] {10.3847/1538-4357/aa9221}, \href
  {https://ui.adsabs.harvard.edu/abs/2017ApJ...850...70T} {850, 70}

\bibitem[\protect\citeauthoryear{{Toomre}}{{Toomre}}{1977}]{Toomre1977}
{Toomre} A.,  1977, in {Tinsley} B.~M.,  {Larson} Richard B.~Gehret D.~C.,
  eds, Evolution of Galaxies and Stellar Populations. p.~401

\bibitem[\protect\citeauthoryear{{Vasiliev}}{{Vasiliev}}{2019}]{vasiliev2019}
{Vasiliev} E.,  2019, \mn@doi [\mnras] {10.1093/mnras/sty2672}, \href
  {https://ui.adsabs.harvard.edu/abs/2019MNRAS.482.1525V} {482, 1525}

\bibitem[\protect\citeauthoryear{{Vasiliev} \& {Athanassoula}}{{Vasiliev} \&
  {Athanassoula}}{2015}]{Vasiliev2015}
{Vasiliev} E.,  {Athanassoula} E.,  2015, \mn@doi [\mnras]
  {10.1093/mnras/stv805}, \href
  {https://ui.adsabs.harvard.edu/abs/2015MNRAS.450.2842V} {450, 2842}

\bibitem[\protect\citeauthoryear{{Vasiliev} \& {Valluri}}{{Vasiliev} \&
  {Valluri}}{2020}]{VasilievValluri2020}
{Vasiliev} E.,  {Valluri} M.,  2020, \mn@doi [\apj] {10.3847/1538-4357/ab5fe0},
  \href {https://ui.adsabs.harvard.edu/abs/2020ApJ...889...39V} {889, 39}

\bibitem[\protect\citeauthoryear{{Vazdekis} et~al.,}{{Vazdekis}
  et~al.}{2015}]{Vazdekis2015}
{Vazdekis} A.,  et~al., 2015, \mn@doi [\mnras] {10.1093/mnras/stv151}, \href
  {https://ui.adsabs.harvard.edu/abs/2015MNRAS.449.1177V} {449, 1177}

\bibitem[\protect\citeauthoryear{{Vogelsberger} et~al.,}{{Vogelsberger}
  et~al.}{2014a}]{Vogelsberger2014a}
{Vogelsberger} M.,  et~al., 2014a, \mn@doi [\mnras] {10.1093/mnras/stu1536},
  \href {https://ui.adsabs.harvard.edu/abs/2014MNRAS.444.1518V} {444, 1518}

\bibitem[\protect\citeauthoryear{{Vogelsberger} et~al.,}{{Vogelsberger}
  et~al.}{2014b}]{Vogelsberger2014b}
{Vogelsberger} M.,  et~al., 2014b, \mn@doi [\nat] {10.1038/nature13316}, \href
  {https://ui.adsabs.harvard.edu/abs/2014Natur.509..177V} {509, 177}

\bibitem[\protect\citeauthoryear{{Weilbacher}, {Streicher}, {Urrutia}, {Jarno},
  {P{\'e}contal-Rousset}, {Bacon}  \& {B{\"o}hm}}{{Weilbacher}
  et~al.}{2012}]{Weilbacher2012}
{Weilbacher} P.~M.,  {Streicher} O.,  {Urrutia} T.,  {Jarno} A.,
  {P{\'e}contal-Rousset} A.,  {Bacon} R.,   {B{\"o}hm} P.,  2012, in
  {Radziwill} N.~M.,  {Chiozzi} G.,  eds,  Society of Photo-Optical
  Instrumentation Engineers (SPIE) Conference Series Vol. 8451, Software and
  Cyberinfrastructure for Astronomy II. p. 84510B, \mn@doi{10.1117/12.925114}

\bibitem[\protect\citeauthoryear{{Weilbacher}, {Streicher}  \&
  {Palsa}}{{Weilbacher} et~al.}{2016}]{Weilbacher2016}
{Weilbacher} P.~M.,  {Streicher} O.,   {Palsa} R.,  2016, {MUSE-DRP: MUSE Data
  Reduction Pipeline} (\mn@eprint {ascl} {1610.004})

\bibitem[\protect\citeauthoryear{{Weinmann}, {van den Bosch}, {Yang}  \&
  {Mo}}{{Weinmann} et~al.}{2006}]{weinmann2006}
{Weinmann} S.~M.,  {van den Bosch} F.~C.,  {Yang} X.,   {Mo} H.~J.,  2006,
  \mn@doi [\mnras] {10.1111/j.1365-2966.2005.09865.x}, \href
  {https://ui.adsabs.harvard.edu/abs/2006MNRAS.366....2W} {366, 2}

\bibitem[\protect\citeauthoryear{{Widrow} \& {Dubinski}}{{Widrow} \&
  {Dubinski}}{2005}]{Widrow2005}
{Widrow} L.~M.,  {Dubinski} J.,  2005, \mn@doi [\apj] {10.1086/432710}, \href
  {https://ui.adsabs.harvard.edu/abs/2005ApJ...631..838W} {631, 838}

\bibitem[\protect\citeauthoryear{{Wilkinson}, {Ludlow}, {Lagos}, {Fall},
  {Schaye}  \& {Obreschkow}}{{Wilkinson} et~al.}{2023}]{Wilkinson2022}
{Wilkinson} M.~J.,  {Ludlow} A.~D.,  {Lagos} C. d.~P.,  {Fall} S.~M.,  {Schaye}
  J.,   {Obreschkow} D.,  2023, \mn@doi [\mnras] {10.1093/mnras/stad055}, \href
  {https://ui.adsabs.harvard.edu/abs/2023MNRAS.519.5942W} {519, 5942}

\bibitem[\protect\citeauthoryear{{Yu}, {Xu}, {Ho}, {Wang}  \& {Kao}}{{Yu}
  et~al.}{2022}]{Yu2022}
{Yu} S.-Y.,  {Xu} D.,  {Ho} L.~C.,  {Wang} J.,   {Kao} W.-B.,  2022, \mn@doi
  [\aap] {10.1051/0004-6361/202142533}, \href
  {https://ui.adsabs.harvard.edu/abs/2022A&A...661A..98Y} {661, A98}

\bibitem[\protect\citeauthoryear{{Yurin} \& {Springel}}{{Yurin} \&
  {Springel}}{2014}]{Yurin2014}
{Yurin} D.,  {Springel} V.,  2014, \mn@doi [\mnras] {10.1093/mnras/stu1421},
  \href {https://ui.adsabs.harvard.edu/abs/2014MNRAS.444...62Y} {444, 62}

\bibitem[\protect\citeauthoryear{{Zana}, {Dotti}, {Capelo}, {Bonoli}, {Haardt},
  {Mayer}  \& {Spinoso}}{{Zana} et~al.}{2018}]{Zana2018}
{Zana} T.,  {Dotti} M.,  {Capelo} P.~R.,  {Bonoli} S.,  {Haardt} F.,  {Mayer}
  L.,   {Spinoso} D.,  2018, \mn@doi [\mnras] {10.1093/mnras/stx2503}, \href
  {https://ui.adsabs.harvard.edu/abs/2018MNRAS.473.2608Z} {473, 2608}

\bibitem[\protect\citeauthoryear{van~den Bosch, {Ferrarese}, {Jaffe}, {Ford}
  \& {O'Connell}}{van~den Bosch et~al.}{1994}]{Bosch1994}
van~den Bosch F.~C.,  {Ferrarese} L.,  {Jaffe} W.,  {Ford} H.~C.,   {O'Connell}
  R.~W.,  1994, \mn@doi [\aj] {10.1086/117179}, \href
  {https://ui.adsabs.harvard.edu/abs/1994AJ....108.1579V} {108, 1579}

\bibitem[\protect\citeauthoryear{van~den Bosch et~al.,}{van~den Bosch
  et~al.}{2007}]{vandenbosch2007}
van~den Bosch F.~C.,  et~al., 2007, \mn@doi [\mnras]
  {10.1111/j.1365-2966.2007.11493.x}, \href
  {https://ui.adsabs.harvard.edu/abs/2007MNRAS.376..841V} {376, 841}

\bibitem[\protect\citeauthoryear{van~den Bosch, {van de Ven}, {Verolme},
  {Cappellari}  \& {de Zeeuw}}{van~den Bosch et~al.}{2008}]{vdBosch2008}
van~den Bosch R.~C.~E.,  {van de Ven} G.,  {Verolme} E.~K.,  {Cappellari} M.,
  {de Zeeuw} P.~T.,  2008, \mn@doi [\mnras] {10.1111/j.1365-2966.2008.12874.x},
  \href {https://ui.adsabs.harvard.edu/abs/2008MNRAS.385..647V} {385, 647}

\bibitem[\protect\citeauthoryear{{van der Kruit} \& {Freeman}}{{van der Kruit}
  \& {Freeman}}{2011}]{Kruit2011}
{van der Kruit} P.~C.,  {Freeman} K.~C.,  2011, \mn@doi [\araa]
  {10.1146/annurev-astro-083109-153241}, \href
  {https://ui.adsabs.harvard.edu/abs/2011ARA&A..49..301V} {49, 301}

\bibitem[\protect\citeauthoryear{{van der Marel} \& {Franx}}{{van der Marel} \&
  {Franx}}{1993}]{vanderMarel1993}
{van der Marel} R.~P.,  {Franx} M.,  1993, \mn@doi [\apj] {10.1086/172534},
  \href {https://ui.adsabs.harvard.edu/abs/1993ApJ...407..525V} {407, 525}

\makeatother
\end{thebibliography}

%\begin{thebibliography}{99}
%\end{thebibliography}

%%%%%%%%%%%%%%%%%%%%%%%%%%%%%%%%%%%%%%%%%%%%%%%%%%

%%%%%%%%%%%%%%%%% APPENDICES %%%%%%%%%%%%%%%%%%%%%

\appendix

\section{Schwarzschild model of FCC170}\label{sec:schw_model}

\begin{figure*}
\includegraphics{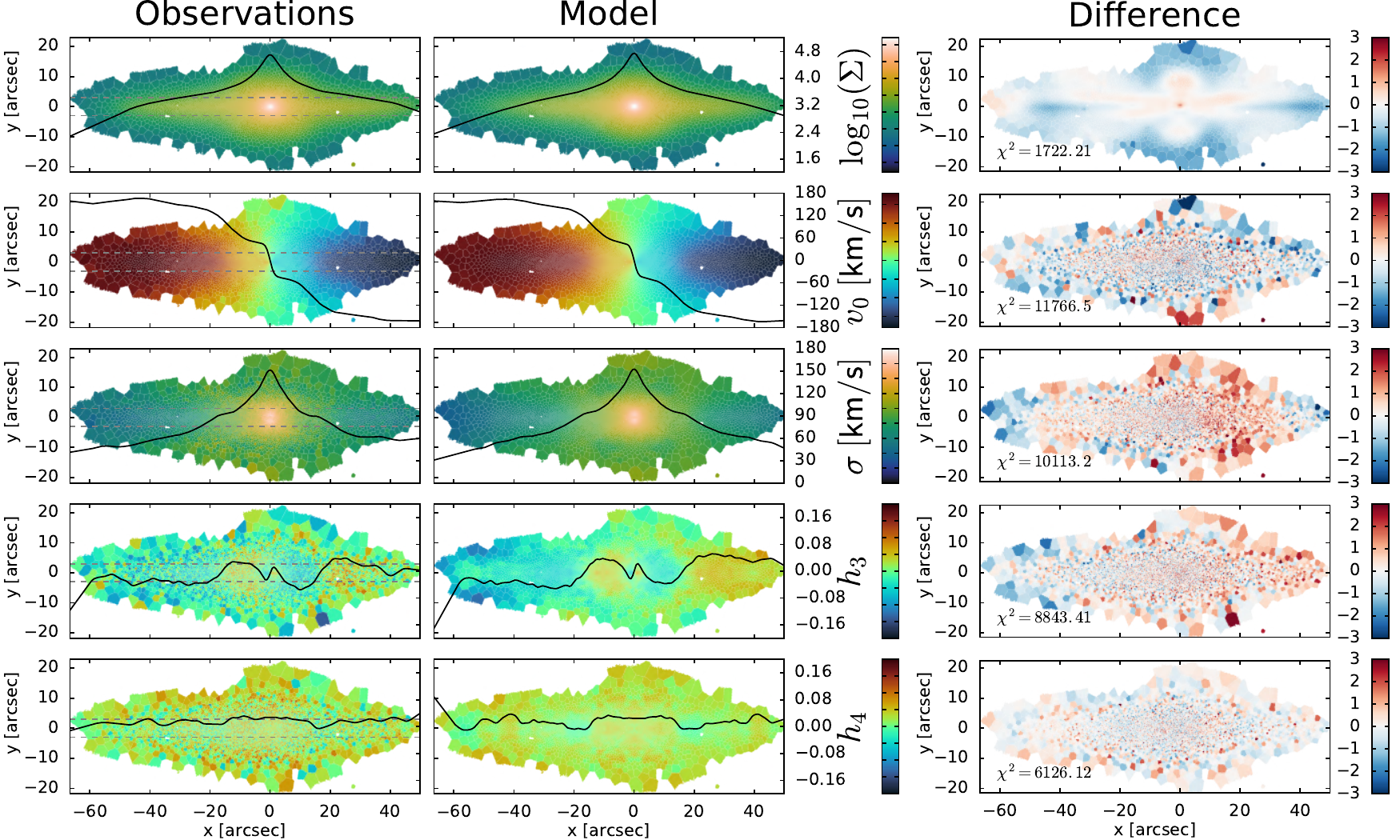}
\caption{Same as Figure~\ref{fig:FCC170_kin_maps}, but for the \citet{Schwarzschild1979} orbit-superposition model of FCC\,170.
} 
\label{fig:FCC170_kin_maps_schw}
\end{figure*}

\begin{figure*}
\includegraphics{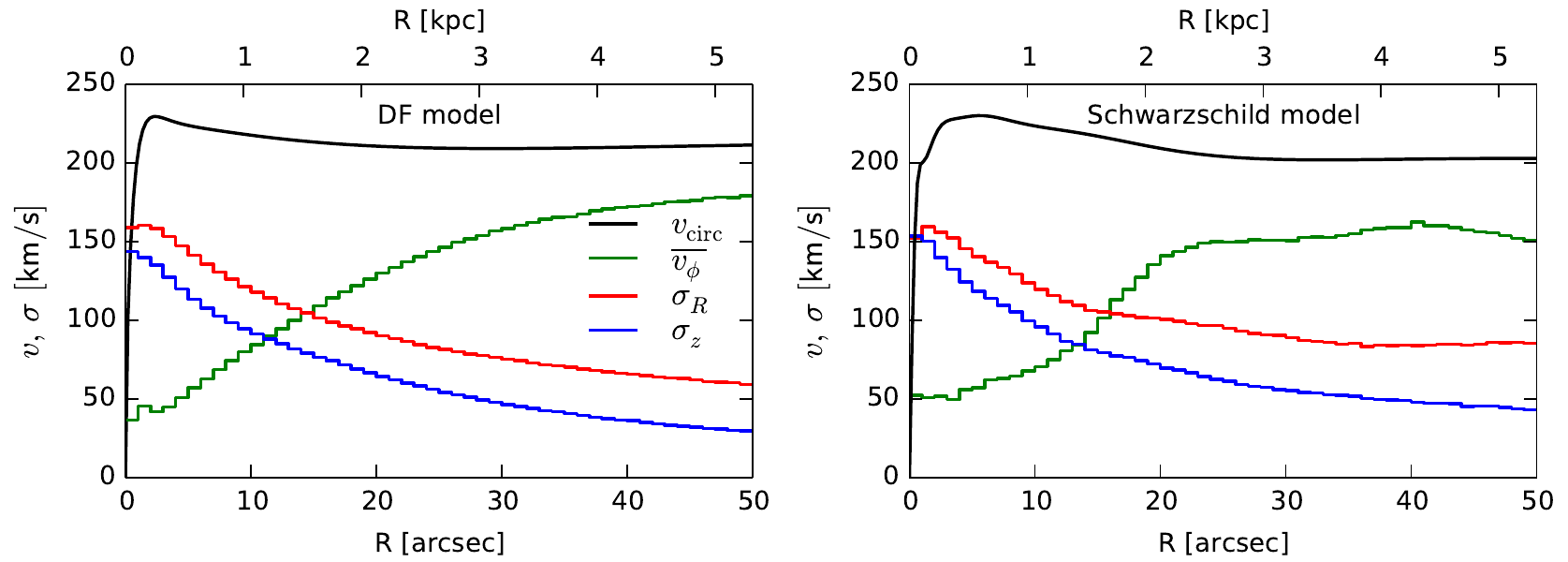}
\caption{Comparison of the internal kinematic properties of the DF-based (left-hand panel) and \citet{Schwarzschild1979} orbit-superposition (right-hand panel) models of FCC\,170. 
The radial profiles of the streaming velocity $\overline{v_\phi}$ (green line), radial velocity dispersion $\sigma_R$ (red line), and vertical velocity dispersion $\sigma_z$ are shown together with the curve of circular velocity $v_{\rm circ}\equiv \sqrt{R\,\mathrm{d}\Phi/\mathrm{d}R}$ (black line) of the total potential.
Both methods produce qualitatively similar profiles, although the orbit-superposition model better matches the bump in the streaming velocity observed at $\sim20$\arcsec and its decline at large radii.
}  
\label{fig:model_kinematics}
\end{figure*}

In addition to the DF-based models described in Section~\ref{sec:df_model}, we also constructed \citet{Schwarzschild1979} orbit-superposition models of the same galaxy, using the \textsc{Forstand} code \citep{VasilievValluri2020}. These models are constrained by the observed kinematic and photometric properties of the galaxy in the form of maps of the Gauss--Hermite moments of the stellar LOSVD and multi-Gaussian parametrization of the surface-brightness distribution. They are similar to the ones presented by \citet{Poci2021}, who relied on another implementation of this method from \citet{vdBosch2008}. Figure~\ref{fig:FCC170_kin_maps_schw} illustrates that these models fit the data largely down to the level of measurement uncertainty, by virtue of having an enormously larger number of (hidden) free parameters -- orbit weights. We use 40000 orbits to fit roughly the same number of observational constraints. The disadvantage of this method is that it does not discriminate \textit{a priori} between different stellar components (i.e. the bulge, NSD, thin and thick discs), which is critical for our application. Although it is possible to assign orbits to these components according to their properties (e.g. circularity), as shown in figure~5 of \citet{Poci2021}, this is beyond the scope of the present study. Figure~\ref{fig:model_kinematics} compares the internal kinematic properties of the DF-based and orbit-superposition models, demonstrating a satisfactory agreement. As our goal in this paper is not to find the best match to the observed galaxy, but rather to construct a suitable approximation with well-defined physical components, we opted to use the multicomponent DF-based models in the rest of our analysis.

\section{Numerical heating of the NSD in an external analytic potential}\label{sec:isolated_NSD}

\placefigNSDanalytic

To illustrate the effect of other galaxy components (primarily the DM halo) on the evolution of the NSD, we conducted the following experiment. We replaced all other components except the NSD and the SMBH by a static \citet{Hernquist1990} potential, whose mass and scale radius are chosen to approximate the circular-velocity curve (equivalently, the cumulative mass profile) of these components combined (primarily the bulge, which is the dominant contribution in the spatial region occupied by the NSD). By doing so, we eliminate the numerical relaxation caused by other components. We then ran an $N$-body simulation of the NSD and the SMBH in \textsc{gizmo} with such external potential.

Figure~\ref{fig:NSD_analytic} shows the snapshots of the self-gravitating NSD component embedded in the external Hernquist potential with a black hole at three different times. As we can observe, the NSD remains unperturbed for the whole run. This suggests that the puffing up of the small razor-thin disc is a consequence of the much larger particle masses in the DM halo, as shown in L19 and L21, and it only occurs when representing the other components as a live $N$-body system. In the right-hand panel, we also plot the circular velocity of the NSD, of the analytic Hernquist profile, and of the bulge, as this last component dominates the central potential up to 1~kpc. Despite the discrepancies at short-kpc scales between the analytic Hernquist and the bulge component, the NSD remains in perfect equilibrium for the entire run and any other choice of parameters for the Hernquist deviates the NSD out from the equilibrium stage.

%%%%%%%%%%%%%%%%%%%%%%%%%%%%%%%%%%%%%%%%%%%%%%%%%%

% Don't change these lines
\bsp	% typesetting comment
\label{lastpage}
\end{document}